\documentclass[aps,showpacs,onecolumn,floatfix,amsmath,amssymb,nofootinbib]{revtex4-1}

\usepackage{amssymb}
\usepackage{amsmath}
\usepackage{epsfig}
\usepackage{graphicx}
\usepackage{color}
\usepackage{latexsym}
\usepackage{bm,latexsym}
\usepackage[mathscr]{eucal}
\usepackage{float}
\usepackage{pbox}

\begin{document}

\title{ Non-linear electrodynamics generalization of the rotating  BTZ black hole. }


\author{Pedro Ca\~nate$^{1}$}
\email{pcanate@fis.cinvestav.mx, pcannate@gmail.com}

\author{Daniela Magos$^{1}$}
\email{dmagos@fis.cinvestav.mx}

\author{Nora Breton$^{1}$}
\email{nora@fis.cinvestav.mx}

\affiliation{$^{1}$Departamento de F\'isica, Centro de Investigaci\'on y de Estudios
Avanzados
del I.P.N.,\\ Apdo. 14-740, Mexico City, Mexico.\\
}

\begin{abstract} 
We obtain two exact solutions of Einstein gravity coupled to nonlinear
electrodynamics (NLED) in $(2+ 1)$-dimensional Anti-de Sitter (AdS) spacetime. 
The solutions are characterized by the mass $M$, angular momentum $J$, cosmological constant or (anti) de Sitter parameter $\Lambda$, and an electromagnetic parameter $Q$, that is related to an electric field in the first solution and to a magnetic charge for the second solution.
Depending on the range of the parameters, the solutions admit a charged rotating asymptotically AdS black
hole (BH) interpretation or a charged rotating
asymptotically AdS traversable wormhole (WH).  If the electromagnetic field is turned
off, the stationary  Ba\~nados-Teitelboim-Zanelli (BTZ) BH is recovered;  in such a way that our BH-WH
solutions are nonlinear charged generalizations of the stationary BTZ-BH.
Moreover, in contrast to the BTZ metric, the derived AdS solutions are singular at
certain radius $r_{s} \neq 0$, resembling the ring singularity of the Kerr-Newman
spacetime; while  if $\Lambda$ is positive the curvature invariants of the second solution are finite.
\end{abstract}

\pacs{04.70.-s, 04.20.Jb, 11.10.Lm}

\maketitle

\section{Introduction}
Asymptotically AdS--BH backgrounds have a great importance because they
provide a valuable theoretical laboratory for studying several classical and quantum
aspects of gravity. In that context, we must emphasize that a variety of physical
phenomena are associated with quantum fields on AdS--BH spacetimes, for
instance,  particle production (Hawking radiation) \cite{Carlip1998}; vacuum
polarization \cite{Sachs2011}, etc.,  Moreover, asymptotically AdS black holes have
recently been of great interest in  relation to the Anti-de Sitter/conformal field
theory (AdS/CFT) correspondence \cite{Carlip2005}. 
Nevertheless, even in the case of free fields the details of such effects are often
difficult to determine.

Thus, as an alternative to understanding the essentials  of the theory (avoiding
unnecessary complications), emerges the study of gravity in $(2+1)$-dimensions,
specifically  due to the fact that General Relativity (GR) in three dimensions is
locally trivial because there are no propagating degrees of freedom. Notwithstanding
this, the relevance of gravity in $(2+1)$-dimensions has been clearly established
since the discovery of the well-known Ba\~nados-Teitelboim-Zanelli (BTZ) stationary
and cyclic symmetry AdS-BH, which 
possesses certain features inherent to $(3 + 1)$-dimensional BHs, and is
characterized by the (anti--)de Sitter parameter $\Lambda$,  mass $M$ and angular momentum $J$
\cite{BTZ1992}.  

The BTZ BH is an exact solution of the three-dimensional Einstein
gravity with cosmological constant, the next logical step is to couple gravity to matter
fields, therefore, over the last years these kind of solutions have been generated.
For instance, the inclusion of Maxwell sources \cite{MTZ}, scalar fields
\cite{scFl},  higher rank tensor fields \cite{HrTf}, gravitational Chern-Simons
terms \cite{CherS}, higher curvature terms \cite{higct}, have been intensively
studied.  

So far stationary Maxwell-electromagnetic  BTZ-BH solutions, have been
obtained by means of  $SL(2,R)$  transformations applied to known electrostatic
solutions. These electromagnetic BTZ solutions are characterized by logarithmic terms in the lapse
function that, if the electromagnetic field is non null, produce divergences at
spatial infinity of some energy--momentum quantities, such that 
asymptotically these metrics do not correspond to the BTZ metric. Some examples of these
solutions are the generalization of the BTZ--BH by
Martinez-Teitelboim-Zanelli (MTZ) \cite{MTZ}, which includes  electric charge $Q$ in
addition to mass $M$, angular momentum $J$ and cosmological constant $\Lambda =-1/l^2$; 
another example is the Clement's spinning charged BTZ solution \cite{Clement}.

In this work we derive electromagnetic  generalizations of the stationary BTZ--BH 
with BTZ asymptotics; we consider  nonlinear electrodynamics (NLED) coupled to
three-dimensional gravity and determine stationary solutions directly by solving the
field equations. 
The derived solutions also can be interpreted as traversable wormholes (WH) in certain
ranges of the parameters. On the other hand, in a stationary cyclic symmetric (SCS) spacetimes the allowed
electromagnetic fields can have electromagnetic invariants that may diverge; we determine the cases 
when it occurs.

The outline of this work is as follows: In Sect. II we introduce the (2+1)-gravity
coupled to NLED, as well as the two formulations of NLED: the one that is in
terms of a Lagrangian that is a function on the electromagnetic  invariant  $F$
($(L,F)$-formalism) and the equivalent formalism in terms of a Hamiltonian that is a function of the
electromagnetic invariant $\mathcal{P}$ ($(\mathcal{H},\mathcal{P})$-formalism); 
we also analyze the finiteness of the electromagnetic invariants 
$F$ and $\mathcal{P}$; then we discuss briefly some of
the previously known  charged stationary BTZ--BH solutions of the Maxwell electrodynamics,
that have been generated from static solutions by using the $SL(2,R)$ symmetry.
In Sect. III we present the first of the  NLED generalization of the rotating BTZ--BH
characterized by the mass $M$, angular momentum $J$, cosmological constant
or (anti) de Sitter parameter $\Lambda$, and an electromagnetic parameter $Q$, and we determine the range of 
the parameters for which the BH or WH interpretation applies; other interesting features of the solutions are the existence of 
an ergoregion, one or two horizons and the BTZ asymptotic limit. In Sect. IV  it is presented the second new solution
that is also a NLED generalization of the stationary BTZ--BH, that is
magnetically charged; as for the first solution, the interpretation of the metric
depends of the range of the parameters,  we roughly analyze its main
characteristics.
For a positive $\Lambda$ the curvature invariants do not diverge, being then a non-singular solution

In the last Section a summary of results and some perspectives of this work are given.
\section{$(2+1)$-dimensional Einstein gravity coupled to nonlinear
electrodynamics}


In this section we present the field equations derived from
the $(2+1)$-dimensional Einstein-NLED action with cosmological constant, 

\begin{equation}\label{actionL}
S[g_{ab},A_{a}] = \int d^{3}x \sqrt{-g} \left( \frac{1}{16\pi}(R - 2\Lambda) +
\frac{1}{4\pi}L  \right),
\end{equation} 
where $R$ is the Ricci scalar, $\Lambda$ is the cosmological constant or de Sitter
parameter, and $L = L(F)$ is a function on the electromagnetic field invariant,
$F=\frac{1}{4}f_{ab}f^{ab}$.

Varying this action with respect to the
gravitational field gives the Einstein equations,

\begin{equation}\label{EinsteinEqsLF}
G_{a}{}^{b} + \Lambda \delta_{a}{}^{b} = 8\pi E_{a}{}^{b},  \quad \textup{ with  } \quad  4\pi E_{a}{}^{b} = L\delta_{a}{}^{b} - L_{F}f_{ac}f^{bc}, 
\end{equation}
where $E_{a}{}^{b}$ is the electromagnetic energy-momentum tensor and $L_{F}$ stands
for the derivative of $L(F)$ with respect to $F$. 
The variation with respect to the electromagnetic potential $A_{a}$, entering in
$f_{ab} = 2\partial_{[a} A_{b]}$, yields the electromagnetic field equations,

\begin{equation}\label{p_NLED}
\nabla_{a}( L_{F} f^{ab}) = 0,
\end{equation}
and the Bianchi identities imply that

\begin{equation}\label{dual_emEqs}
\nabla_{a}(_{\ast}\!\boldsymbol{f} )^{a b} = 0,
\end{equation}
where $(_{\ast}\!\boldsymbol{f} )^{a}$ denotes the Hodge star operation (or
Hodge dual) with respect to the metric.

Alternatively, one can define the anti-symmetric tensor $P_{a b} = L_{F}
f_{a b}$ \cite{Pleban}, then (\ref{p_NLED}) becomes $\nabla_{a}( P^{ab}) = 0$ and
the electromagnetic  description is now in terms of the Hamiltonian function,
$\mathcal{H}$, obtained from the Lagrangian by means of a Legendre transformation

\begin{equation}\label{actionH}
\mathcal{H} = 2F L_{F} -  L. 
\end{equation} 
 It can be shown that $\mathcal{H}$ is a function depending on the electromagnetic
invariant $\mathcal{P}$, defined as $\mathcal{P} = \frac{1}{4}P_{ab}P^{ab} = (L_{F})^{2}F$. Moreover, (by denoting the derivative of $\mathcal{H}$ with respect to $\mathcal{P}$, as $\mathcal{H}_{\mathcal{P}}$) it can be shown that $\mathcal{H}_{\mathcal{P}}=1/L_{F}$, 
 
\begin{equation}
\frac{d\mathcal{H} }{dF} = L_{F} + 2FL_{FF}  = \frac{ 1 }{ L_{F} } \frac{d
}{dF}\left( (L_{F})^{2}F \right) = \frac{ 1 }{ L_{F} } \frac{d\mathcal{P}}{dF} \quad
\Rightarrow \quad \left( \frac{d\mathcal{H} }{dF} \right)
\left(\frac{d\mathcal{P}}{dF} \right)^{-1}  = \mathcal{H}_{\mathcal{P}} = \frac{ 1
}{ L_{F} }.
\end{equation}
Therefore, by using the previous relation and $ \mathcal{P} = (L_{F})^{2}F$, the
nonlinear electromagnetic Lagrangian, $L(F)$, in terms of $\mathcal{P}$ and
$\mathcal{H}$, is given by 
$L = 2\mathcal{P}\mathcal{H}_{\mathcal{P}} - \mathcal{H}$. Now, by using
$L=L(\mathcal{P})$, $L_{F}=L_{F}(\mathcal{P})$, and the tensor $P_{a b}$, the
field equations (\ref{EinsteinEqsLF}) can be written as, 

\begin{eqnarray}
&&G_{a}{}^{b} + \Lambda \delta_{a}{}^{b} = 8\pi E_{a}{}^{b}, \quad \textup{ with  }
\quad  4\pi E_{a}{}^{b} = - \mathcal{H}_{\mathcal{P}}P_{ac}P^{bc}  + 
(2\mathcal{P}\mathcal{H}_{\mathcal{P}} - \mathcal{H})\delta_{a}{}^{b}, 
\end{eqnarray}
Whereas the electromagnetic field equations (\ref{p_NLED}) and (\ref{dual_emEqs})
become 

\begin{equation}\label{emEqs}
\nabla_{a}(P^{ab}) = 0 = \nabla_{a} [  ( _{\ast}\boldsymbol{P} )^{a} \mathcal{H}_{P}],
\end{equation}
where $( _{\ast}\boldsymbol{P})_{n}$, in terms of the components $P^{ab}$,  is given
by 

\begin{equation}
( _{\ast}\boldsymbol{P} )_{n} = \frac{ \sqrt{ -g } }{ 3 } \left( P^{tr}
\delta_{n}{}^{\phi} + P^{r\phi} \delta_{n}{}^{t} + P^{\phi t} \delta_{n}{}^{r} 
\right), \quad n = t, r, \phi.
\end{equation}

The nonvanishing components of the electromagnetic field should be in agreement with the spacetime symmetries; in 
the next subsection we determine the allowed electromagnetic fields in stationary metrics with cyclic symmetry. 
 
\subsection{Classification of NLED fields in stationary cyclic symmetric spacetimes} 
 
It turns out that the form of the electromagnetic field $P_{a b}$ admitted by the
SCS spacetimes is not arbitrary, and there is a theorem that classifies the nonlinear electromagnetic fields, $f_{ab}$,
admitted by these kind of spacetimes \cite{CB2018}, that in coordinates $\{$ $t$, $r$, $\phi$ $\}$ can be written as

\begin{equation}\label{StCyc}
ds^{2} = g_{tt}(r)dt^{2} + 2g_{t\phi}(r)dt d\phi + g_{\phi \phi}(r)d \phi^{2} +
g_{rr}(r)dr^{2}. 
\end{equation}

Through the constitutive
relation $P_{ab} = L_{F} f_{ab}$ the theorem can be easily formulated for
$P_{ab}$, in the following way.

{\it\textbf{Theorem 1.}} The general form of  the stationary cyclic symmetric
electromagnetic fields in $(2+1)$-dimensions in general relativity coupled to
nonlinear electrodynamics, with the electromagnetic field
$(P^{ab}, \mathcal{H})$  given by,

\begin{equation}\label{Ps_ab}
P^{\alpha\beta} = L_{F}f^{\alpha\beta} = \frac{ 1 }{\sqrt{-g}}  \left[ 
\begin{array}{ccc}
0 & b & -\frac{3g_{rr}c}{\sqrt{-g}}L_{F} \\
-b & 0 & a \\
\frac{3g_{rr}c}{\sqrt{-g}}L_{F} & - a & 0
\end{array} \right] 
= \frac{ 1 }{\sqrt{-g}}  \left[  \begin{array}{ccc}
0 & b & -\frac{3g_{rr}}{\sqrt{-g}}\frac{c}{\mathcal{H}_{P}} \\
-b & 0 & a \\
\frac{3g_{rr}}{\sqrt{-g}}\frac{c}{\mathcal{H}_{P}} & - a & 0
\end{array} \right].
\end{equation}
Or in terms of its dual $_{\ast}\boldsymbol{P}$, as,

\begin{equation}\label{Theorem1}
_{\ast} \boldsymbol{P} = \frac{g_{rr}}{\sqrt{-g}} \frac{c}{\mathcal{H}_{P}} dr + 
\frac{a}{3}dt + \frac{b}{3}d\phi,
\end{equation} 
where $a$, $b$ and $c$ are constants, that by virtue of the Ricci circularity
conditions, are constrained to the cases 

\begin{equation}\label{Theorem1a}
ac = 0 = bc, 
\end{equation}
that gives rise to two disjoint branches or classes, the first being (class 1)

\begin{equation}\label{Theorem1b}
c \neq 0,\quad a = 0 =b \quad \Rightarrow \quad _{\ast} \boldsymbol{P} =
\frac{g_{rr} }{\sqrt{-g}}\frac{c}{\mathcal{H}_{P}}dr,  
\end{equation}
while the second branch is (class 2),

\begin{equation}\label{Theorem1c}
c = 0 \quad \Rightarrow \quad _{\ast} \boldsymbol{P} = \frac{a}{3}dr +
\frac{b}{3}d\phi,
\end{equation}
with its own sub-classes; subclass 2.a with $a \ne  0$ and $b=0$; and sub-class 2.b with $b \ne 0$ and $a=0$.

If the SCS metrics can be interpreted as BHs, then the electromagnetic invariants can diverge at the BH horizon, in the following subsection we analyze when this occurs. 

\subsection{ Behavior of the electromagnetic invariants $\mathcal{P}$ and  $F$ in
stationary cyclic symmetric spacetimes}

Let us consider the general form for SCS (2+1) spacetimes in coordinates $\{$ $t$, $r$, $\phi$ $\}$, 

\begin{equation}\label{Scyclic}
ds^{2} = -N^{2}(r)dt^{2} + \frac{dr^{2}}{H^{2}(r)} + r^{2}(d\phi + \omega(r)dt)^{2},
\end{equation}
where $N(r)$, $H(r)$ and $\omega(r)$ are functions depending only on the radial
coordinate. In these metrics the nonvanishing electromagnetic field components,
$P_{ab}$ calculated from the matrix (\ref{Ps_ab}), are,

\begin{equation}
P_{tr} = \frac{ -N^{2}b + (b\omega - a)r^{2}\omega }{rHN}, \quad  P_{t\phi} =
\frac{3c}{ \mathcal{H}_{P} }, \quad  P_{r\phi} = \frac{(a- b\omega)r}{HN},  
\end{equation}
or in  terms of the orthonormal tetrad (see Appendix I)  $P_{(a)(b)}$ are, 
 
\begin{equation}
P_{(0)(1)} = -\frac{b}{r},  \quad P_{(0)(2)} = \frac{3c}{rN \mathcal{H}_{P} }, \quad P_{(1)(2)} = \frac{a-b\omega}{N}.   
\end{equation}

Specifically for the branch $c\neq0$, $a=0=b$ or class 1, it is obtained that $\boldsymbol{P} =
P_{t\phi} dt \wedge d\phi =  P_{(0)(2)} \theta^{(0)}\wedge \theta^{(2)}$; these
solutions are called hybrid in \cite{GarciaBook}. 

For the electromagnetic invariants, calculated using Eq. (\ref{Ps_ab}),
the following considerations apply:

\begin{enumerate}

\item In the class 1 $c \neq 0$, the invariants $F$ and $\mathcal{P}$ are given by,

\begin{equation}\label{F_P_c}
F = - \frac{9 c^{2} }{r^{2}N^{2} } \quad \Rightarrow \quad \mathcal{P} = - \frac{9
c^{2} }{r^{2}N^{2} \mathcal{H}^{2}_{\mathcal{P}} }     
\end{equation}
If the metric (\ref{Scyclic}) allows a black hole interpretation, then the  event
horizons $r_{h}$ will be identified as the zeros of the function $N(r)$, i.e.,
$N(r_{h}) =0.$  This means that for NLED in the branch $c \neq 0$, the construction of
black holes with invariant $\mathcal{P}$ regular at $r_{h}$ is possible only if
$\mathcal{H}_{\mathcal{P}}(r_{h}) N(r_{h})$ does not vanish at $r_{h}$; necessarily
the invariant $F$ will not be well defined at $r= r_{h}$.  Specifically, for
Maxwell-electrodynamics $F = \mathcal{P}$ and according to Eq.  (\ref{F_P_c}), in
this branch there no exist three-dimensional charged rotating black holes with well
defined electromagnetic invariants at $r_{h}$.

\item In the class 2 $a \neq 0 \neq b$, the invariants $F$ and $\mathcal{P}$ are
given by

\begin{equation}\label{2invariants}
\mathcal{P} = -\frac{b^{2}}{r^{2}} + \frac{(a-b\omega)^{2}}{N^{2}} \quad \Rightarrow
\quad F = -\frac{ b^{2}\mathcal{H}^{2}_{P} }{r^{2}} + \frac{(a-b\omega)^{2}
\mathcal{H}^{2}_{P} }{N^{2}}.      
\end{equation}
Now, specifically when  $( a - b \omega(r_{h}) )/N(r_{h})$  and 
$\mathcal{H}^{2}_{P}(r_{h})$ are finite, then the construction of
black holes with $F$ and $\mathcal{P}$ well defined at $r = r_{h}$ is possible.

While in the sub-class 2.a  $a \neq 0=b$, only if $\mathcal{H}_{\mathcal{P}}(r_{h}) /
N(r_{h})$ is finite,  it is possible to construct BHs with well defined  invariant
$F$ at $r=r_{h}$, but $\mathcal{P}$ will be singular there.  In the sub-class 2.b  $b \neq 0=a$, 
only if $\mathcal{H}_{\mathcal{P}}(r_{h})$ and $ \omega (r_h) /
N(r_{h})$ are finite,  it is possible to construct BHs with well defined  invariants
$F$ and $\mathcal{P}$  at $r=r_{h}$.
\end{enumerate} 

In this paper we present two solutions, the first one  corresponding to  class 1 (analyzed in the next section),
characterized by the invariants in Eq. ( \ref{F_P_c} ) and such that
$\mathcal{P}(r_h)$ is finite but $F$ diverges.
The second solution  belongs to sub-class 2.a (section V), and
is such that, when the solution admits black hole interpretation,
$F(r_h)$ is finite at the horizon but $\mathcal{P}(r_h)$ diverges.

Approaching nonlinear electrodynamics it is under debate which one of the electromagnetic fields 
$f_{ab}$ or $P_{ab}$ is the most physically significant. We quote the opinion by Plebanski 
in \cite{Pleban}, page 21: {\it We will consistently accept the point of view that $P_{ab}$  
is fundamental and $f_{ab}$ the secondary object.}; then regarding the divergence of $F$ or 
$\mathcal{P}$, sticking to this view, it is the first of our solutions the most physically interesting, with 
$F(r_h)$ divergent and  $\mathcal{P}(r_h)$ finite at the horizon (in the cases where the solution 
admits a BH interpretation).

\subsection{Charged generalizations of the rotating BTZ in Maxwell electrodynamics}

In the linear limit or  Maxwell electrodynamics, i.e., $L(F) = F=P= \mathcal{H}(P)$, 
one charged rotating generalization of the BTZ black hole is the
Martinez-Teitelboim-Zanelli (MTZ) solution \cite{MTZ}, that includes, besides the
AdS parameter $l$, the mass $M$, an electric charge $Q$ and the angular velocity of the boost $\omega_0$. 
This solution was obtained from the static one ($\omega_0=0$), given by

\begin{equation}
ds^{2} = -\left(- M + \frac{r^{2}}{l^{2}} - \frac{Q^{2}}{4}\ln r^{2} \right)dt^{2} +
\frac{dr^{2}}{ - M + \frac{r^{2}}{l^{2}} - \frac{Q^{2}}{4}\ln r^{2} } +
r^{2}d\phi^{2};    
\end{equation}
by the application of the $SL(2, R)$ symmetry, through the Lorentz boost 

\begin{equation}
\tilde{t} \rightarrow \frac{t - \omega_0 \phi}{\sqrt{ 1 - \frac{\omega_0^{2}}{ l^{2} }
}}, \quad \tilde{\phi} = \frac{ \phi - \frac{ \omega_0 }{ l^{2} } t }{\sqrt{ 1 -
\frac{\omega_0^{2}}{ l^{2} } }},
\label{SL_transf}
\end{equation}
it is obtained the MTZ solution,  

\begin{eqnarray}
&& ds^{2} =  - \left[ \frac{r^{2}}{l^{2}} -  \frac{ l^{2} }{ l^{2} - \omega_0^{2}
}Z(r) \right] dt^{2} + \left[ r^{2} +  \frac{ l^{2} \omega_0^{2}  }{ l^{2} -
\omega_0^{2} }Z(r) \right] d\phi^{2}  + \frac{ dr^{2}}{\frac{r^{2}}{l^{2}} - Z(r)
} - \frac{ 2l^{2} \omega_0 }{ l^{2} -  \omega_0^{2} }Z(r) dt d\phi, \label{MTZ} \\ 
&&   
Z(r) = \left( M + \frac{Q^{2}}{4}\ln r^{2} \right), \quad  \boldsymbol{A} = -\frac{
Q \ln r }{2\sqrt{ 1 - \omega_0^{2}/l^{2} } } (dt - l^{2} \omega_0 d\phi ).  
\end{eqnarray}

Another example of a rotating solution obtained by means of the coordinate transformation
(\ref{SL_transf}) is the  Clement's spinning charged BTZ solution \cite{Clement},
given by 

\begin{eqnarray}
ds^{2} &=& - \left( W(r)  - \frac{\omega_0^{2}}{l^{4}}r^{2} \right) dt^{2} + \left(
r^{2} - \omega_0^{2} W(r) \right) d\phi^{2} - 2 \omega_0
Q^{2}\ln\left(\frac{r^{2}}{r_{0}^{2}}\right) dt d\phi + \frac{dr^{2}}{ W(r)  }
\label{Clement} \\ 
W(r)&=& \frac{r^{2}}{l^{2}} - Q^{2}\ln\left(\frac{r^{2}}{r_{0}^{2}}\right), \quad
\boldsymbol{A} = \frac{Q}{2}\ln\!\left(\frac{r^{2}}{r_{0}^{2}} \right) (dt - \omega_0
d\phi).    
\end{eqnarray}
Both solutions (\ref{MTZ}) and (\ref{Clement}) allow a black hole interpretation.

In the context of nonlinear electrodynamics in (2+1)-gravity with cosmological constant there is a solution in the subclass 2.b ($a=0, b \ne 0$) derived by Cataldo and Garcia \cite{Cataldo}; this is a SCS solution of the form (\ref{Scyclic}) with

\begin{eqnarray}
 N^2(r) & =& H^2(r)= -M- \Lambda r^2- q^2 \ln (r^2+a^2), \quad \omega(r)=0, \nonumber\\
 L_{NLED}(r) &=& \frac{q^2}{8 \pi} \frac{r^2-a^2}{(r^2+a^2)^2}, \quad E(r)= \frac{q r^3}{(r^2+a^2)^2},
\end{eqnarray}
where $q$ is an electromagnetic charge and $L_{NLED}(r)$ is the NLED field; in case $q=0$ the static BTZ--BH is recovered. Alike the previous mentioned solutions,
the presence of logarithmic terms in the metric components 
lead to divergences on the energy--momentum quantities at infinity and therefore
the metric asymptotics are different from the BTZ one.

In contrast, the NLED generalization of the rotating BTZ black hole that we present in the next section
is asymptotically BTZ.
\section{NLED generalization of the rotating BTZ black hole} \label{Sec2}

\begin{figure}[!ht]\centering
\includegraphics[scale=0.6]{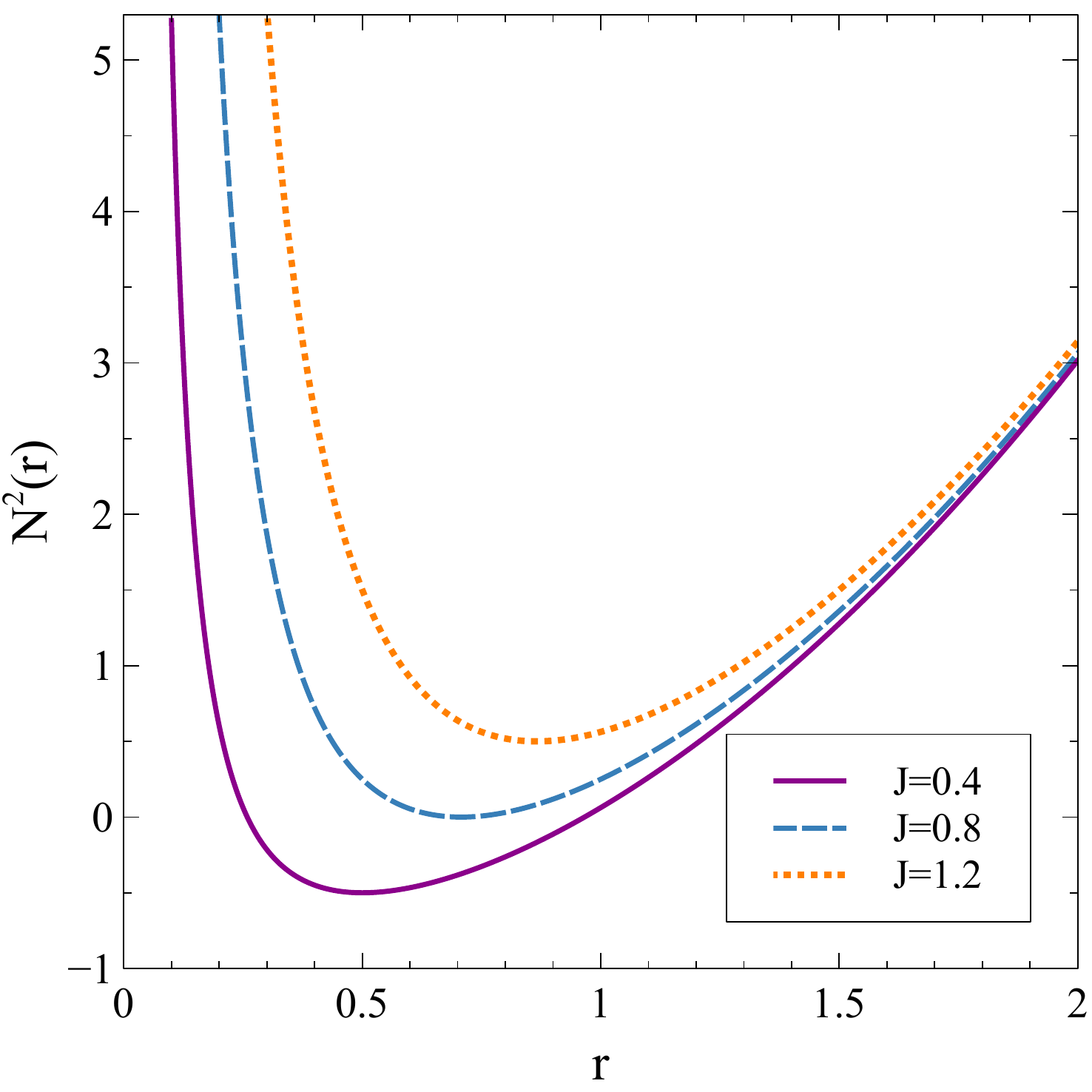}
\caption{\small The lapse function $N^2(r)$ is shown for different values of $J$, and fixed values
of mass $M=1$ and $\Lambda=1.5$. Notice that if $J>1$ the function has no real
zeros.}\label{N2}
\end{figure}

The gravitational field of the derived hybrid $(c\neq0)$ stationary solution  is given by the
metric 

\begin{equation}\label{generalSol}
ds^{2} = - N^{2}(r) dt^{2}  + \frac{(2Mr^{2} - J^{2})^{2} }{ S^{2}(r)  N^{2}(r)} dr^{2} + r^{2}\left[d\phi + \left(\frac{ S(r)  }{ 2Jr^{2} } - \frac{ \sqrt{ M^{2} - c^{\lambda} } }{ J } \right)dt\right]^{\!2},  
\end{equation}
with 
\begin{equation}\label{H_model_sol}
N^{2}(r) = - M - \Lambda r^{2} + \frac{ J^{2} }{ 4r^{2} }, \quad S^{2}(r) = J^{4} - 4 J^{2} M r^{2} + 4(M^{2} - c^{\lambda})r^{4}, \quad \omega(r) =  \frac{ S(r) }{ 2Jr^{2} }  - \frac{ \sqrt{ M^{2} - c^{\lambda}}}{J}.      
\end{equation}
The solution is characterized by the mass $M$, angular momentum $J$, cosmological constant
$\Lambda$ that can be positive (de Sitter) or negative (anti-de Sitter), and the electromagnetic parameter $c^{\lambda}$ that can be positive or negative. Note that the lapse function $N(r)$ is the same as for the stationary BTZ--BH, but the rest of the metric functions are higher order polynomials. In figure (\ref{N2}) the metric function $N^2(r)$ is shown for different values of $J$.
The specific function $\mathcal{H}$, determining the nonlinear electrodynamics
source, is given by

\begin{equation}\label{H_model}  
\mathcal{H} = \mathcal{H}(\mathcal{P}(r)) = \frac{ c^{\lambda} \left( 4 \Lambda M
r^{6} - 6 J^{2}\Lambda r^{4} - 3J^{2}Mr^{2} + \frac{J^{4}}{2} \right) }{( -2Mr^{2} +
J^{2})^{3}} \quad \textup{ with } 1<\lambda<2,
\end{equation}

The electromagnetic invariant $\mathcal{P}(r) $ and 
$\mathcal{H}_{\mathcal{P}} = \mathcal{H}_{,r}/\mathcal{P}_{,r}$ are given, respectively, by 

\begin{equation}\label{P_r_1}
\mathcal{P}(r) = -\frac{1}{2} \left( \frac{3c}{ r N \mathcal{H}_{\mathcal{P}} } 
\right)^{2}  = -\frac{ J^{4} c^{2\lambda - 2} }{ 72 } \frac{ (-4\Lambda r^{4} - 4 M
r^{2} + J^{2} )^{3} }{ ( -2Mr^{2} + J^{2})^{6} },
\end{equation}

\begin{equation}\label{H_P}
\mathcal{H}_{\mathcal{P}} =  -\frac{ 36 c^{ 2 - \lambda } ( -2Mr^{2} + J^{2})^{3} }{
J^{2} (-4\Lambda r^{4} - 4 M r^{2} + J^{2} )^{2}  }.
\end{equation}
Whereas the invariant $F$ is $F = \mathcal{H}^{2}_{\mathcal{P}} \mathcal{P} = -9
c^2/(2r^2 N^2(r))$. At spatial infinity the nonvanishing electromagnetic field component goes 
to a constant, $P_{(0)(2)} (r \mapsto \infty)= J^2c^{ \lambda -1}/(6 M^3l^3).$

The corresponding curvature invariants are given by 

\begin{eqnarray}
R &=& \frac{ \mathcal{A}_{1}(r) }{ ( -2Mr^{2} + J^{2})^{3} }, \label{R_es}\\
R_{\alpha\beta}R^{\alpha\beta} &=& \frac{ \mathcal{A}_{2}(r) }{ ( -2Mr^{2} +
J^{2})^{6} } ,\label{Ric_es}\\
R_{\alpha\beta\mu\nu}R^{\alpha\beta\mu\nu} &=& \frac{ \mathcal{A}_{3}(r) }{ (
-2Mr^{2} + J^{2})^{6} }, \label{Riem_es}
\end{eqnarray}
being $\mathcal{A}_{n}(r)$, with $n = 1, 2, 3$, polynomial functions on $r$, that do not cancel 
the factors $(-2Mr^{2} + J^{2})$ in the denominator. The
region determined by $r = r_{s} = J/ \sqrt{ 2M }$, corresponds  then to a curvature
singularity since the invariants, (\ref{R_es}), (\ref{Ric_es}), (\ref{Riem_es}), blow up at $r = r_{s}$.  

\subsection{BTZ--limit cases}

The BTZ limit is obtained in two cases: 
(i) by turning off the electromagnetic field we
get the stationary BTZ-BH; and (ii) when $J \mapsto 0$ that corresponds to a static BTZ-BH.
Moreover, at infinity $r \mapsto \infty$, we recover the static BTZ--BH.
In what follows we give the details of these limits.


\begin{itemize}

\item{\bf Limit $c \rightarrow 0$, with $J\neq0$ }

For $c = 0$, the electromagnetic field is turned off, $P_{ab}=0$. Moreover, in this case the function $S^{2}(r)$ becomes 
$S^{2}(r) = J^{4} - 4 J^{2} M r^{2} + 4 M^{2} r^{4} = ( 2Mr^{2}- J^{2} )^{2}$ , and then the line element (\ref{H_model_sol}) takes the form, 

\begin{eqnarray}\label{BTZ}
ds^{2} = &-& \left( - M - \Lambda r^{2} + \frac{ J^{2} }{ 4r^{2} }  \right)dt^{2}  + \frac{ dr^{2} }{ \left( - M - \Lambda r^{2} + \frac{ J^{2} }{ 4r^{2} }  \right)}  +  r^{2}\left[d\phi + \left( \frac{ 2Mr^{2} - J^{2}  }{ 2Jr^{2} } - \frac{ M }{ J } \right) dt\right]^{\!2} \\
= &-& \left( - M - \Lambda r^{2} + \frac{ J^{2} }{ 4r^{2} }  \right)dt^{2}  + \frac{ dr^{2} }{ \left( - M - \Lambda r^{2} + \frac{ J^{2} }{ 4r^{2} }  \right)} + r^{2}\left(d\phi - \frac{J}{ 2r^{2} } dt\right)^{\!\!\!2},
\end{eqnarray}
which (for $M>0$, and $\Lambda = -1/l^{2}$, i.e. $\Lambda <0$) corresponds to the rotating BTZ black hole solution. 
\item {\bf Limit $J \rightarrow 0$, with $c\neq0$ } 
In this case it is obtained that,

\begin{equation}
\lim_{J \rightarrow 0}S^{2}(r) =  4(M^{2} - c^{\lambda})r^{4}, \quad \lim_{J \rightarrow 0} \left(\frac{ S(r)  }{ 2Jr^{2} } - \frac{ \sqrt{ M^{2} - c^{\lambda} } }{ J } \right) = 0, \quad \lim_{J \rightarrow 0} N^{2}(r) = - M - \Lambda r^{2},   
\end{equation}
then, the line element (\ref{generalSol}), takes the form, 

\begin{equation}\label{J_0_generalSol}
ds^{2} = - \left( - M - \Lambda r^{2} \right) dt^{2}  + \frac{M^{2}}{(M^{2} - c^{\lambda}) \left( - M - \Lambda r^{2} \right) } dr^{2} + r^{2}d\phi^{2}.
\end{equation}
That by renaming 

\begin{equation}\label{M_tilde_L_tilde}
\tilde{M} = \frac{(M^{2} - c^{\lambda})}{M}, \quad \tilde{\Lambda} = \frac{(M^{2} - c^{\lambda})\Lambda}{M^{2}}, 
\end{equation}
can be written as

\begin{equation}\label{J_0_tilde_generalSol}
ds^{2} = - \left( - \tilde{M} - \tilde{\Lambda} r^{2} \right) d\tilde{t}^{2}  + \frac{1}{\left( - \tilde{M} - \tilde{\Lambda }r^{2} \right) } dr^{2} + r^{2}d\phi^{2}, \quad \textup{ with } \quad \tilde{t} = \frac{M^{2}}{(M^{2} - c^{\lambda})} t.
\end{equation}

On the other hand, in this case the quantities $\mathcal{H}$ of (\ref{H_model}), and $\mathcal{P}\mathcal{H}_{\mathcal{P}}$ of (\ref{P_r_1}) and (\ref{H_P}), become  

\begin{equation} 
\mathcal{H} = \mathcal{H}(\mathcal{P}(r)) = - \frac{ c^{\lambda} \Lambda }{ 2 M^{2} }, \quad \mathcal{P}\mathcal{H}_{\mathcal{P}} = 0,
\end{equation}
  
Then the action (\ref{actionL}), with $L = 2\mathcal{P}\mathcal{H}_{\mathcal{P}} - \mathcal{H}$, becomes,

\begin{equation}
S[g_{ab},A_{a}] = \int d^{3}x \sqrt{-g} \left( \frac{1}{16\pi}(R - 2\Lambda) + \frac{1}{4\pi}\frac{ c^{\lambda} \Lambda }{ 2 M^{2} }  \right) =
\int d^{3}x \sqrt{-g} \left( \frac{R - 2\tilde{\Lambda}}{16\pi} \right),
\end{equation} 
and the metric (\ref{J_0_tilde_generalSol}) for $ \tilde{M}>0$, and $\tilde{\Lambda} = -1/\tilde{l}^{2}$, (i.e. $\tilde{\Lambda} <0$) corresponds to the static BTZ black hole solution.

\item{\bf BTZ Asymptotic behavior $r \mapsto \infty$ }

In order to show that at infinity the solution is the static BTZ, we determine the asymptotic form of the metric  at  $r \mapsto \infty$, in this limit the metric components behave like,
\begin{equation}
\lim_{r \rightarrow  \infty}\frac{(2Mr^2-J^2)^2}{S^{2}(r)} =  \frac{M^2}{(M^{2} - c^{\lambda})}, \quad
\lim_{r \rightarrow \infty} \left(\frac{ S(r)  }{ 2Jr^{2} } - \frac{ \sqrt{ M^{2} - c^{\lambda} } }{ J } \right) = 0, \quad \lim_{r \rightarrow \infty} N^{2}(r) = - M - \Lambda r^{2},   
\end{equation}
then, the line element (\ref{generalSol}), takes the same  form as Eq. (\ref{J_0_generalSol}), and therefore by renaming to $\tilde{M}, \quad \tilde{\Lambda}$ and $\tilde{t}$ the metric acquires the static BTZ form, Eq. (\ref{J_0_tilde_generalSol}).

\end{itemize}

\subsection{Behavior of the solution }

In this subsection we address the different cases arising from the nature of the
roots of the metric functions $N(r)$ and $S(r)$, i.e. the interpretation of the solution depends if the
roots are real or complex.\\
We shall consider that $\Lambda = -1/l^{2}$ in order that the solution
(\ref{generalSol})--(\ref{H_model_sol}) has the AdS asymptotics; also that $(M^{2} - c^{\lambda}) >0$. \\
To clarify the analysis of the roots, we shall use the re-scaling as in \cite{Cruz1994},
that will be denoted with a hat,
\begin{equation}\label{hats1}
\hat{t}= \frac{t \sqrt{M}}{l}, \quad  \hat{r}= \frac{r}{l \sqrt{M}}, \quad \hat{\phi}=\sqrt{M} \phi, \quad \hat{J}=\frac{J}{lM}, 
\quad \hat{Q}=\frac{M^2-c^\lambda}{M^2}.
\end{equation}
The line element is then

\begin{eqnarray}\label{H_model_sol_BH}
ds^2=&-&\left(\frac{\hat{J}^2}{4\hat{r}^2}+ \hat r^2- 1\right)d\hat{t}^2+
\frac{ (\hat{J}^2-2\hat r^2)^2}{({\hat{J}^2}/{4\hat{r}^2}+ \hat r^2- 1)(\hat
J^4-4\hat J^2 \hat r^2 + 4\hat Q \hat r^4)} d\hat r^2\nonumber\\
 &+&\hat r ^2 \left[d\hat \phi + \left(\frac{\sqrt{\hat J^4-4\hat J^2 \hat r^2 +
4\hat Q \hat r^4}}{2\hat J \hat r^2} -\frac{\sqrt{\hat Q}}{\hat J}\right)d\hat
t\right]^2.
\end{eqnarray}

Thus, the roots of the functions $N^{2}(r)$ and $S^{2}(x)$ defined in
(\ref{H_model_sol}) are given, respectively in terms of $\hat{Q}$ and  $\hat{J}$, given in Eq. (\ref{hats1}), by,

\begin{equation}\label{HW_r0}
\hat {r}^2_{\pm}=\frac{1}{2}\left(1\pm\sqrt{1-\hat{J}^2}\right); \quad
\hat{x}^2_{\pm}=\frac{\hat{J}^2}{2\hat{Q}}\left(1 \pm \sqrt{1-\hat{Q}}\right),
\end{equation}
where $\hat{r}_\pm$ are the roots of $N^2(r)$ and $\hat{x}_{\pm}$ are the roots of
$S^2(r)$. 

Clearly for all $\hat{J}$ and $\hat{Q}$ $\in \mathbb{R}$ such that $\hat{r}_{\pm}$ and
$\hat{x}_{\pm}$ are real numbers, it is fulfilled that,   

\begin{equation}\label{radios}
(\hat{r}_{-})^{2} \leq \hat{r}_{s}^{2} \leq (\hat{r}_{+})^{2} \quad \textup{ and }
\quad  (\hat{x}_{-})^{2} \leq \hat{r}_{s}^{2} \leq (\hat{x}_{+})^{2},
\end{equation}
where the singularity radius is $\hat{r}_s=\frac{\hat{J}}{\sqrt{2}}$.
The roots $(\hat{r}_{-})^{2}$ and $(\hat{r}_{+})^{2}$, of $N(r)$, coincide only when
$\hat{J} =1$. In this case we have that $(\hat{r}_{-})^{2} = (\hat{r}_{+})^{2} =
\hat{r}_{s}^{2} = 1/{ 2 }$. Similarly, the roots $(\hat{x}_{-})^{2}$
and $(\hat{x}_{+})^{2}$, are equal if $\hat{Q}=1$, implying that $(\hat{x}_{-})^{2} =
(\hat{x}_{+})^{2} = \hat{r}_{s}^{2} = {\hat{J}^{2}}/{ 2 }$. \\

The solution presents different features depending on the range of the parameters 
$\hat{Q}$ and   $\hat{J}$, that we classify in the following cases,

\begin{enumerate}
\item{\bf BH with one event horizon at $\hat{r}_{h} = \hat{r}_{+}$:  Case
$\hat{r}_{+}\in\mathbb{R}^{+}$ and  $\hat{x}_{+}\in\mathbb{C}$.} 

By restricting to $\hat{Q}>1$, from Eq. (\ref{HW_r0}), it is obtained that 
$S(r)$ has no real roots, i.e. 
$S^{2}(r)\in
\left\{ \mathbb{R}^{+} - \{ 0\} \right\}$, for all $r$ and $\hat{x}_{\pm}\in\mathbb{C}$.  If, additionally,
we restrict to  $\hat{J}\leq 1$, then the two roots of $N(r)$, ${N}^{2}(\hat{r}_{\pm})=0$, are real.
According to Eq. (\ref{radios}), in $(2+1)$-dimensions, the region $\hat{r}=\hat r_{-}$ is completely enclosed by the singularity at $\hat{r} = \hat r_{s}$; therefore the region $\hat{r}=\hat r_{-}$ cannot be reached by any observer. 

It turns out that there is a stationary limit,  the component $g_{tt}$ of the metric
(\ref{H_model_sol_BH}), vanishes at $\hat{r}=\hat{r}_{\rm erg}$, given by

\begin{equation}
\hat{r}^2_{\rm erg}=\frac{2 \hat{Q}+\sqrt{\hat{Q} (\hat{J}^4+4 \hat{Q}-4 \hat{J}^2
\hat{Q})}}{4\hat{Q}-\hat{J}^2}\;>\hat r_+ .\label{rerg}
\end{equation}
The region at $\hat{r}_{\rm erg}$ is then an ergoregion that lies outside of the horizon.

If the electromagnetic field is turned off, $c=0\Rightarrow \hat{Q}=1$, then, the
{\it ergoradius} reduces to $\hat{r}_{\rm erg}^2 =1$ or $r_{\rm erg}=l^2 M$
that consistently corresponds to the ergoregion of the BTZ black hole, since when $\hat{Q}=1$ the BTZ solution is recovered.

In figure (\ref{RootsA}) we can see that the region $\hat{r}=\hat r_{s}$ is completely
enclosed by the region $\hat{r}=\hat r_{+}$, and the following inequality is met 
\begin{equation}
\hat{r}_s<\hat{r}_h=\hat{r}_+<\hat{r}_{\rm erg}. \nonumber
\end{equation}

Hence, for the setting of parameters $\{$ $\Lambda = -1/l^{2}$, $\hat{Q}>1$, $
\hat{J}\leq 1$  $\}$, the metric (\ref{H_model_sol_BH}) has the structure of a
$(2+1)$-dimensional charged rotating AdS black hole, with event horizon at $\hat{r}= \hat{r}_{h}=
\sqrt{\frac{1}{2}\left(1+\sqrt{1-\hat J^2}\right)}$, and  curvature singularity at
$\hat{r} = \hat r_{s} = \frac{\hat J^{2}}{ 2 }$, in this way generalizing the rotating BTZ black hole; this is illustrated 
in Fig. \ref{RootsA}.

\begin{figure}[H]\centering
\includegraphics[scale=0.52]{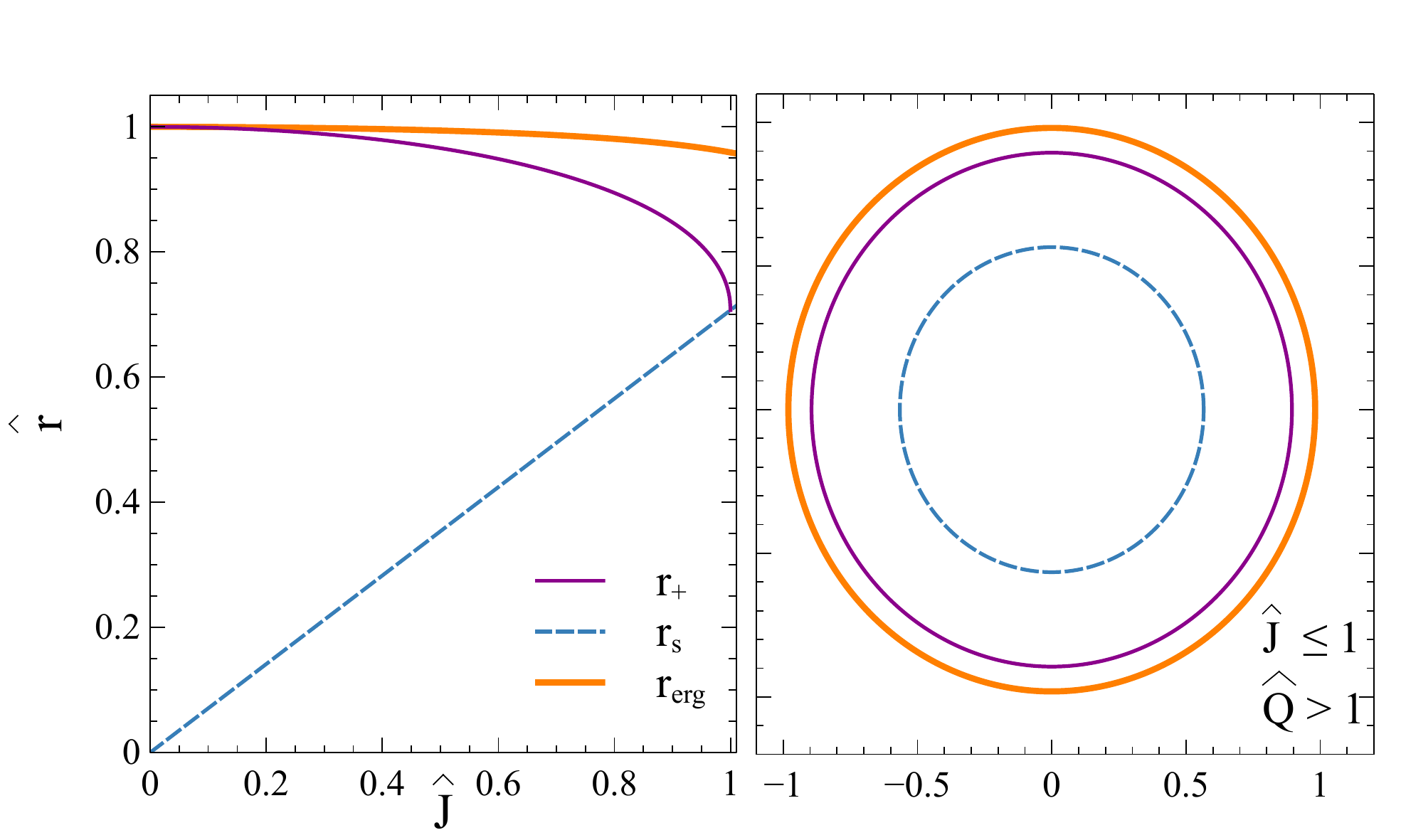}
\caption{\small One horizon BH case is illustrated.
To the left $\hat{Q}$=1.2. The dashed curve represents the
curvature singularity $\hat{r}_s$, the purple curve is the event horizon
$\hat{r}_+$ and the orange one (exterior curve) is the ergoregion radius
$\hat{r}_{\rm erg}$. The root $\hat r_-$ (not shown in the plot) is inside $\hat r_s$.}
\label{RootsA}
\end{figure}


\item{\bf BH with two horizons: Case $\hat r_+ $ $\in \mathbb R^+$ and $\hat x_+$ $\in \mathbb R^+$.}
 
According to  Eqs. (\ref{HW_r0}), by assuming that $\hat{J} \le 1$ and $\hat{Q} \le 1$  then the roots are real, and in this case there are two horizons and one ergoregion, this is depicted  in figure (\ref{RootsB}).
 
There are two regions of interest, the first region 
$\hat{r}_s<\hat{x}_+ <\hat{r}_+<\hat{r}_{\rm erg}$, where the inner horizon
$\hat{r}_h^{in}$ is  $ \hat{r}_h^{in}= \hat x_+$ while the outer horizon $\hat{r}_h^{out}$ is $\hat{r}_h^{out}=\hat{r}_+$. In the second region there is an interchange on the position of the horizons, such that $\hat{r}_s<\hat{r}_+<\hat{x}_+<\hat{r}_{\rm erg}$.
\begin{figure}[H]\centering
\includegraphics[scale=0.5]{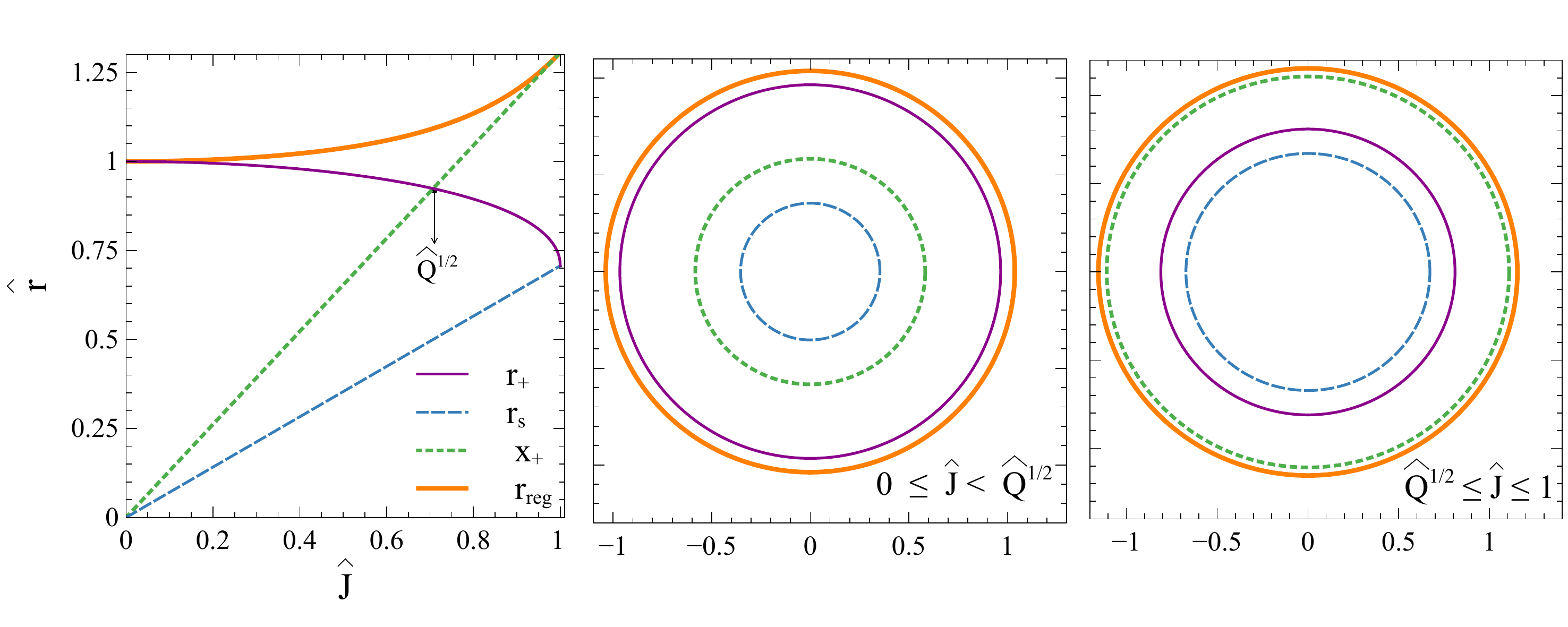}
\caption{\small The two horizon  BH case is shown. The parameters $\hat{Q}$ and $\hat{J}$ are both less than 1; to the left $\hat{Q}=0.5$. The dashed curve represent the curvature singularity $\hat{r}_s$, the dotted one is $\hat{x}_+$, the purple curve is $\hat{r}_+$ and the exterior curve
represents the ergoregion $\hat{r}_{\rm erg}$.
In  the  graphic in the middle $\hat{x}_+$ is the inner horizon and $\hat{r}_+$ is the
outer one. In the graphic to the right the positions of the horizons are switched.}\label{RootsB}
\end{figure}

\item[\bf 2b.] {\bf  BH by complex extension. Case $\hat r_{+} = \hat{x}_{+}$.}

In this case we have a BH with one horizon, with the roots of $N(r)$ and $S(r)$ being real and such that $\hat r_{+} = \hat{x}_{+}$. This case occurs
if $\hat {J}=\sqrt{\hat Q}$. Substituting into Eq. (\ref{rerg}) it gives the
ergoregion radious $\hat{r}_{\rm erg}$, 
\begin{equation}
 \hat r_{\rm erg}^2= \frac{2}{3}\left( 1+\sqrt{1-\frac{3}{4}\hat J^2}\right),
\end{equation}

such a region exists if $\hat J^2\leq 4/3$, that is always true since $\hat
J^2 \leq 1$. Also it is fulfilled that $\hat r_+=\hat x_+ < \hat r_{\rm erg}$ as we can
see in figure (\ref{Bex}).
\begin{figure}[H]\centering
\includegraphics[scale=0.5]{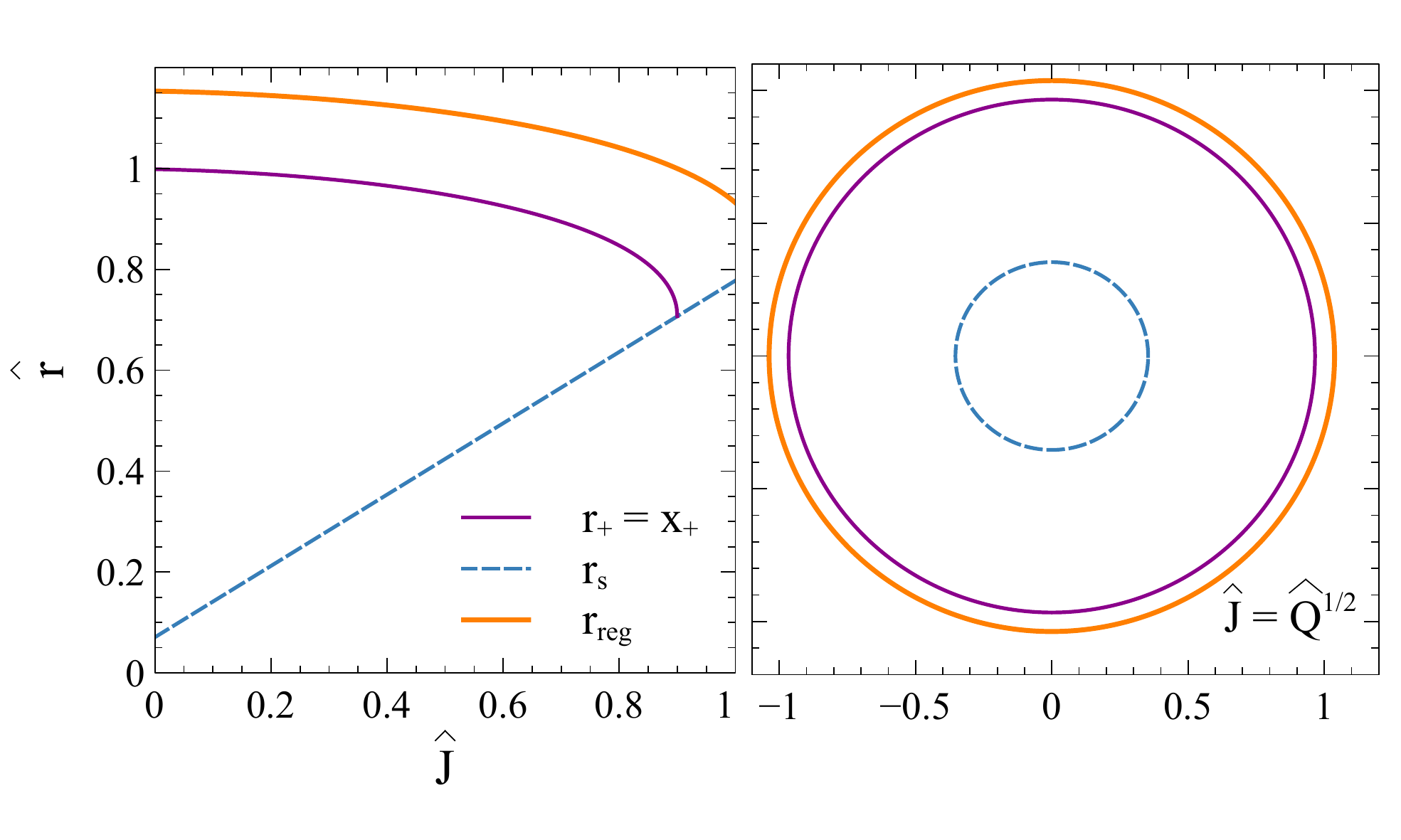}
\caption{\small The  case  where the roots of $S(r)$ and $N(r)$ coincide, $\hat r_+ =\hat x_+$ for $\hat{J}=\sqrt{\hat{Q}}$.  }\label{Bex}
\end{figure}
In this case the line element (\ref{H_model_sol_BH}) takes the form

\begin{eqnarray}\label{extremoA}
 ds^2=-\left(\frac{\hat{J}^2}{4\hat r^2}+\hat r^2-1\right)d\hat{t}^2 +
\frac{(\hat{J}^2-2\hat r^2)^2}{ 4\hat J^2\hat r^2\left(\frac{\hat{J}^2}{4\hat
r^2}+\hat r^2-1\right)^2} d\hat r^2\nonumber\\
 +\hat r ^2\left[ d\hat \phi +\left( \frac{1}{\hat r}\left(\frac{\hat{J}^2}{4\hat
r^2}+\hat r^2-1\right)^{1/2}-1\right)d\hat t\right]^2.
\end{eqnarray}
which is valid for the domain $\hat r\geq  \hat r_h=\hat r_+=\hat x_+$.

However in the region $\hat r_{s} \leq \hat r< \hat r_+$, it turns out that $N^{2}(r)
\leq 0$; this implies that the line element changes its signature and is a complex
metric.

Now, by the complex transformation $\tilde{t} = i\hat t$, $\tilde{\phi} = \hat \phi
- \hat t$, the metric becomes real, this is
\begin{equation}
 ds^2=-\left(1-\hat r^2-\frac{\hat{J}^2}{4\hat r^2}\right)d\tilde{t}^2
 +  \frac{(\hat{J}^2-2\hat r^2)^2}{ 4\hat J^2\hat r^2\left(1-\hat
r^2-\frac{\hat{J}^2}{4\hat r^2}\right)^2} d\hat r^2
 +\hat r ^2\left[ d\tilde \phi+\left( \frac{1}{\hat r}\sqrt{\left(1-\hat
r^2-\frac{\hat{J}^2}{4\hat r^2}\right)}\right)d\tilde t\right]^2.
\end{equation}

one can see that the signature of the metric is the same in both regions, in the
inner region $(\hat r_{s} \leq \hat r\leq \hat r_+)$ while in the outer region $(\hat r_+
\leq \hat r\leq \infty)$, being $\hat r_+=\hat x_+$ the extreme black hole eventBH horizon.

\end{enumerate}

\subsection{Geodesic equations}

We briefly sketch the main features of geodesics for massive and massless test particles in the neighborhood of the BH.
In a static and axisymmetric spacetime, the energy $E$ and the angular momentum
$L_z$ of a test particle of mass $m$ and momentum $P^i =m\frac{dx^i}{d\tau}$, with
$x^i=t, r,\phi$ and $\tau$ being the affine parameter along the geodesics, are conserved quantities,
\begin{eqnarray}\label{conserved}
 E\,&=&-P_t = (N^2(r)-r^2 \omega^2)m\dot t -r^2\omega m \dot\phi, \nonumber\\
L_z&=&P_\phi= r^2\omega m \dot t+r^2 m \dot\phi,
\end{eqnarray}
where the functions $N^2(r)$ and $\omega(r)$ are defined in equations
(\ref{H_model_sol}) and the dot represents the derivative with respect the
parameter $\tau$.\\

The geodesic equations are 
\begin{eqnarray}
\dot t &=&\;\frac{1}{N^2(r)}(E+\omega L_z),\label{dot_t}\\
\dot \phi &=& \;\frac{1}{N^2(r)}\left(\frac{N^2(r)}{r^2}L_z-\omega ( E + \omega L_z ) \right),\label{dot_phi}\\
\frac{N^2(r)}{H^2(r)}\;\dot r^2 \;&=&\; (E+\omega L_z)^2
+N^2(r)\left(\frac{L_z}{r^2}-\alpha\right), \label{dot_r}
\end{eqnarray}
where from now on $E$ and $L_z$ are the energy and angular momentum per unit of mass. The constant $\alpha$ is determined by the scalar product
$g^{ij}\frac{dx^i}{d\tau} \frac{dx^j}{d\tau}=\alpha$, and can take the values
$\{$-1, 0, 1 $\}$, corresponding to time, null and space-like geodesics respectively.

\begin{figure}[H]\centering
\includegraphics[scale=0.6]{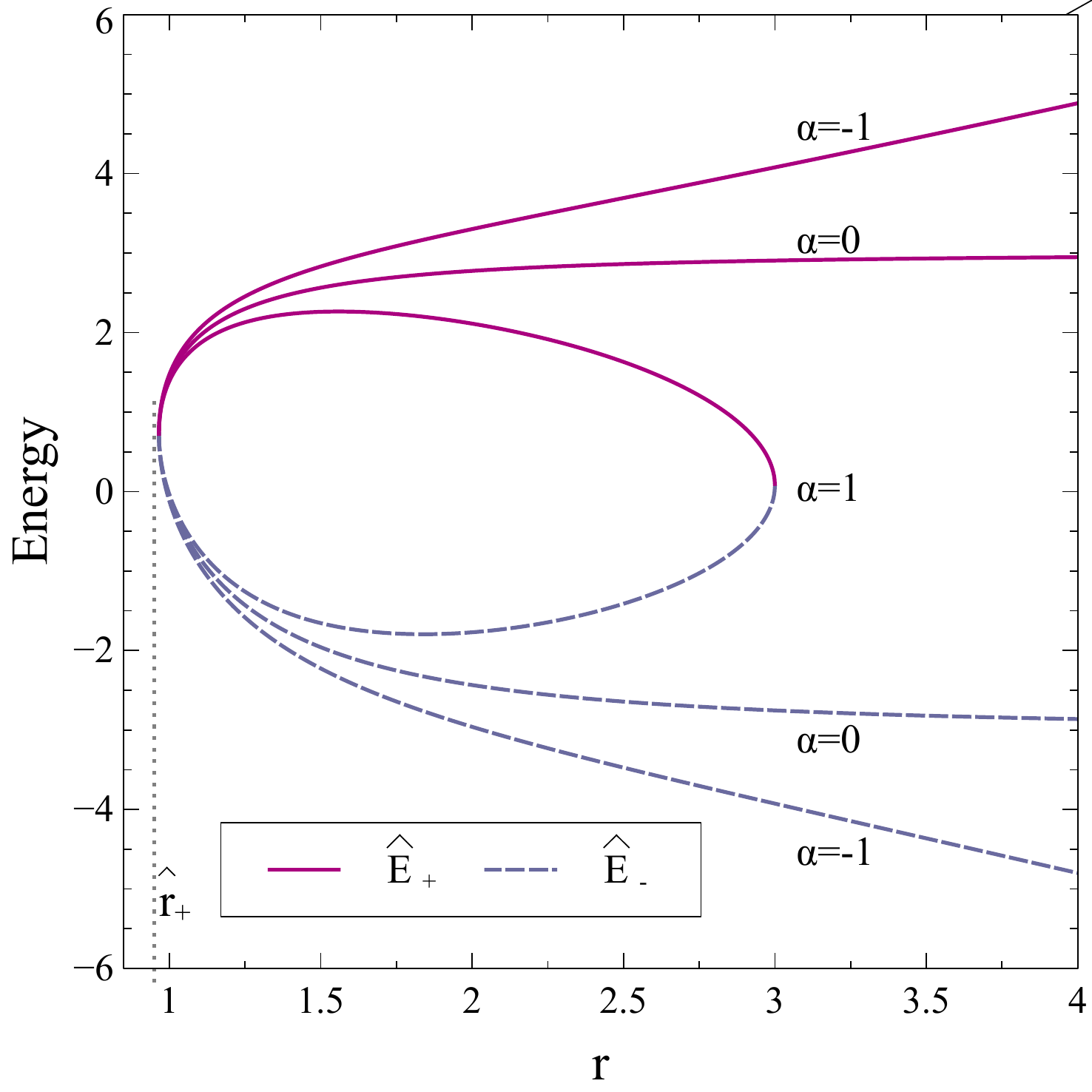}
\caption{\small The energy of test particles at the turning points in the potential of the BH. We observe regions with negative energy then the Penrose energy extraction mechanism would be possible.  In this plot $\hat{J}=0.5,\,\hat{L}_z=3, \hat{Q}=1.2$; $\alpha = -1, 0,  1,$ for time, null and space-like geodesics respectively.}\label{EnA}
\end{figure}

We solve the radial equation (\ref{dot_r}) for $E$ at the turning points $\dot r =0$,
and write down the solution in  terms of the hat variables given in (\ref{hats1}) 

\begin{equation}\label{energy}
 \hat E_\pm=-\hat\omega \hat L_z\pm \frac{1}{\hat r}\sqrt{\hat N^2(\hat r)(\hat
L_z^2-\alpha \hat r^2),}
\end{equation}

where we are defining
\begin{equation}\label{hats2}
\hat E=\frac{E}{\sqrt{M}}, \hspace{1.2cm} \hat L_z
=\frac{L_z}{l\sqrt{M}},\hspace{1.2cm}\hat\omega= l\omega, \hspace{1.2cm}\hat
N^2(\hat r)=\frac{N^2(r)}{M}.
 \end{equation}
In  Fig. (\ref{EnA}) is shown the solution (\ref{energy}) for
the case $\hat{r}_+ \in \mathbb{R}^+,\; \hat{x}_+\in\mathbb{C}$, the other cases have a similar behaviour: massive particles are always trapped in the BH, but massless not necessarily; the regions between the plots are forbidden. Negative energy states exist in the ergoregion, that, in principle, enables the Penrose process by massless particles. 

\begin{figure}[H]\centering
\includegraphics[scale=0.6]{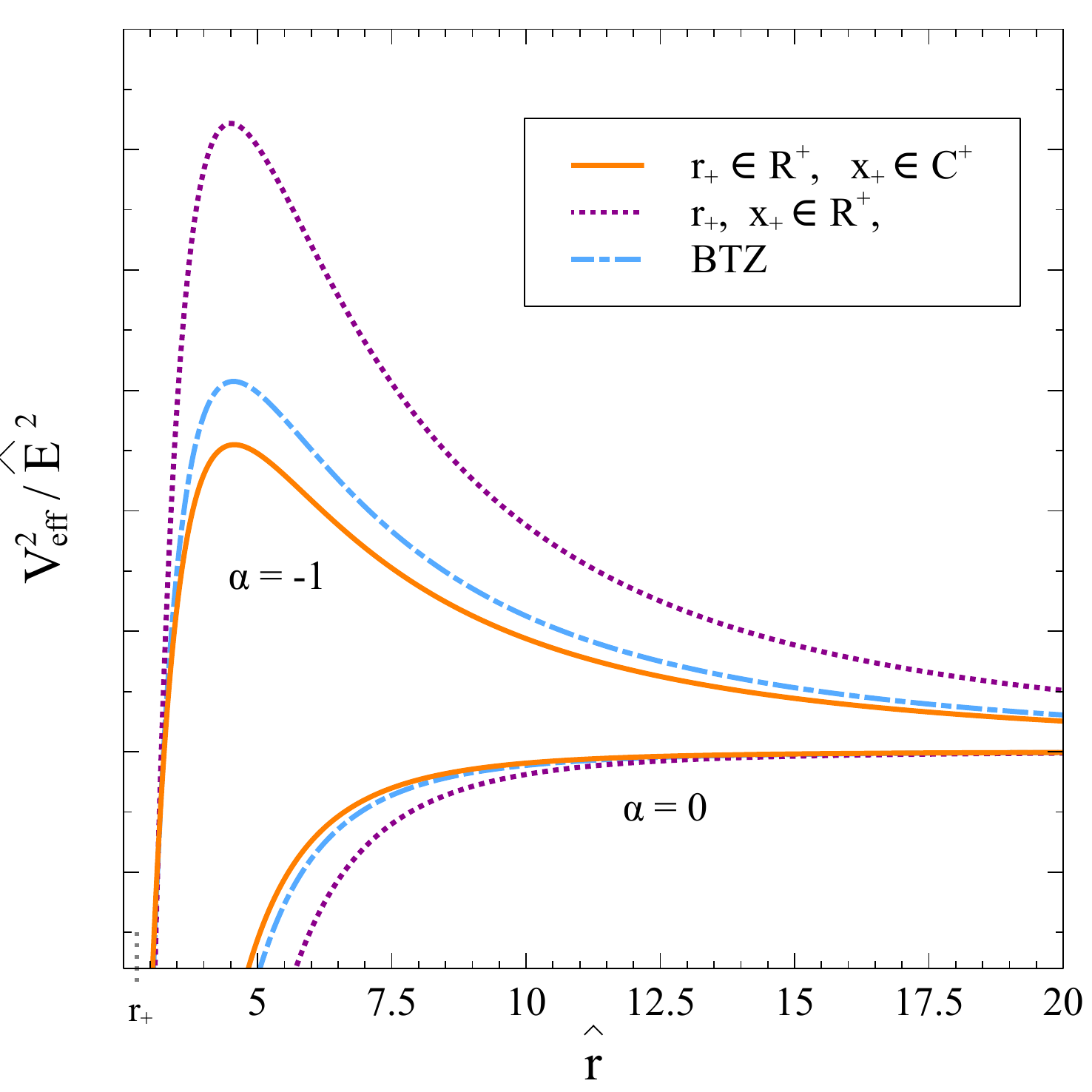}
\caption{\small The effective potential that feels a test particle in the BH spacetime, compared with the BTZ effective potential; the effective potentials asymptotically tend to the BTZ one. $\alpha=0$ is for massless particles while $\alpha=-1$ corresponds to massive particles. The values of the parameters are  $\hat{J}=0.5,\; \hat{E}=5,\; \hat{L_z}=3$, and the charge is varied as
{\it Solid curve:} $\hat{Q}=1.2$, {\it dotted curve:} $\hat{Q}=0.6$, {\it BTZ
(dashed):} $\hat{Q}=1$ .}\label{VeffT}
\end{figure}

Rewriting the radial equation (\ref{dot_r}), we have

\begin{equation}
 \hat{\dot r}^2 =-\hat{E}^2\left(\hat{N^2}(\hat{r})\frac{\hat{L_z}^2}{\hat{r}^2
\hat{E}^2}-\left(1+\hat{\omega}\frac{\hat{L_z}}{\hat{E}}\right)^2-\alpha \frac{\hat
N^2(\hat r)}{\hat E^2}\right)\frac{4\hat{J}^2
\hat{r}^4\hat{\omega}^2}{(\hat{J}^2-2\hat{r}^2)^2},
\end{equation}
and using $\hat{\dot r}^2= \hat{E}^2-V_{\rm eff}^2$, we can identify the effective
potential term as

\begin{equation}
 V_{\rm eff}^2=\hat{E}^2\left[1+\left(\hat{N}^2(\hat{r})\frac{\hat{L_z}^2}{\hat{r}^2
\hat{E}^2}-\left(1+\hat{\omega}\frac{\hat{L_z}}{\hat{E}}\right)^2 -\alpha
\frac{\hat N^2(\hat r)}{\hat E^2}\right)\frac{4\hat{J}^2
\hat{r}^4\hat{\omega}^2}{(\hat{J}^2-2\hat{r}^2)^2}\right], \label{Veff}
\end{equation}
that is depicted in Fig. \ref{VeffT} jointly with the BTZ effective potential.

From Fig. \ref{VeffT} we see that for timelike geodesics (massive test particles) there is a barrier with unstable circular geodesics at its maximum; while for larger $r$ it is monotonically decreasing, showing then the absence of stable circular orbits, very similar to the BTZ effective potential. For the null geodesics (massless particles) the effective potential is monotonically decreasing as the particle approaches the BH, i.e. the behavior is Schwarzschild-like. More details on the BTZ--BH geodesics may be consulted in \cite{Cruz1994}.

\subsection{ Wormhole Interpretation }\label{whLimit}

Due to the extraordinary connection between entanglement and wormholes recently
suggested by Maldacena and Susskind \cite{MaldacenaSusskind}, the interest in
wormhole geometries have been renewed.

In this subsection we shall determine the setting of the parameters in order that the solution
(\ref{H_model_sol_BH}) admits a traversable wormhole (WH) interpretation.

The general metric of a $(2+1)$-dimensional SCS-WH  is given by 

\begin{equation}\label{whAnsatz}
ds^{2}_{wh} = -e^{2\Phi(r)} dt^{2} + \frac{dr^{2}}{ 1 - \frac{b(r)}{r}}  + r^{2} [
d\phi + \omega(r)dt ]^{2}, 
\end{equation}
where, according to  \cite{ThorneMorris}, $\Phi(r)$ and $b(r)$ are functions of the
radial coordinate $r$. $\Phi(r)$ is called the red-shift function, for it is
related to the gravitational redshift; whereas $b(r)$ is the shape function.  The
radial coordinate has a range that increases from the minimum value $r_{0}$ (WH throat 
$b(r_{0}) = r_{0}$) up  to $r \rightarrow \infty$.

On the other hand, for the WH to be traversable, one must demand the absence
of event horizons, which are identified as the regions where $e^{2\Phi(r)} = 0$;  if $e^{2\Phi(r)}$ is a continuous, nonvanishing and
finite function in the whole range of $r$,  $r\in[r_{0},\infty)$, then there are no horizons. 

Besides, for a WH be traversable it is required  the fulfilment of the flaring out
condition (deduced from the mathematics of embedding),  given by 
\begin{equation}
\frac{b(r) - rb(r)_{,r}}{b^{2}(r)} > 0.\label{trav_cond}
\end{equation}

Note that at the throat  $b(r_{0}) = r_{0},$ and  the flaring out condition reduces
to $b_{,r}(r_{0})< 1$. Connecting with the energy momentum tensor $E_{(\alpha)(\beta)}$ through the Einstein 
field equations, the flaring out
condition is equivalent to the violation of the Null Energy Condition (NEC), which
establishes that $E_{(\alpha)(\beta)}n^{(\alpha)}n^{(\beta)}\geq0$  for a null
vector $n^{(\alpha)}$ [see Appendix II for details]. To preserve the signature of the metric the
condition $(1 - b/r) \geq 0$ is also imposed.

Having established these WH generalities, we will show that for certain
setting of the parameters the general solution (\ref{H_model_sol_BH}) represents a 
traversable $(2+1)$-dimensional charged and rotating WH.

By comparison between the line elements (\ref{Scyclic}) and (\ref{whAnsatz}), we
obtain, 

\begin{equation}\label{WH_fun}
e^{2\Phi(r)} = N^{2}(r), \;\quad 1-\frac{b(r)}{r} = H^2(r)= \frac{N^2(r) S^2(r)}{(2Mr^2-J^2)^2},\; \quad  \omega(r) =
\frac{ S(r)  }{ 2Jr^{2} } - \frac{ \sqrt{M^{2} - c^{\lambda}} }{ J },
\end{equation}
with the metric functions given by ($\ref{H_model_sol}$).\\
Recalling that the roots of $N^2(r)$ are given in Eqs. (\ref{HW_r0}),
$\hat r_{\pm}^{2} = \frac{1}{2} \left(1\pm \sqrt{1-\hat J^2}\right)$, then by
demanding that $\hat{J}>1$, it turns out that the roots are complex,  i.e. $N(r)=0$ has no real roots; then $e^{2\Phi(r)}
= N^{2}(r)\in \left\{ \mathbb{R}^{+} - \{ 0\} \right\}$, for all $r$, and no horizons occur.

If additionally it is imposed that $\hat{Q}<1$ then one can define a domain, $r \in
[\hat{r}_{0}, \infty)$ with $\hat{r}_{0}$ given by  $H^2(\hat{r}_0)=0$, i.e, $\hat{r}_0$ is the WH throat,
\begin{equation}
\hat{r}_{0}^2=\hat{x}_{+}^2 =\frac{\hat{J}^2}{2 \hat{Q}}\left(1+\sqrt{1-\hat{Q}}\right), 
\end{equation}
such that for all $r\in (\hat{r}_{0}, \infty)$ it is fulfilled that $\omega(r) > 0$,
$1-b(r)/r > 0$ and  $b(r_{0}) = r_{0}$.

Moreover the WH is traversable because the flaring out
condition is fulfilled, i.e.  $b'(r_{0})<1$. By using $b(r)$ in (\ref{trav_cond}), one
finds that
\begin{eqnarray}
-\frac{2(\hat{J}^2-\hat{Q})}{\sqrt{1-\hat{Q}}}&<&0, \quad 
\Rightarrow\hspace{1cm} \hat{J}^2>  \hat{Q},
\end{eqnarray} 
that is always true because we have assumed $\hat{J} > 1$ and  $Q \leq 1$. Thus, the flaring out condition is satisfied and therefore the metric  admits a traversable WH interpretation.

\begin{figure}[H]\centering
\includegraphics[scale=0.6]{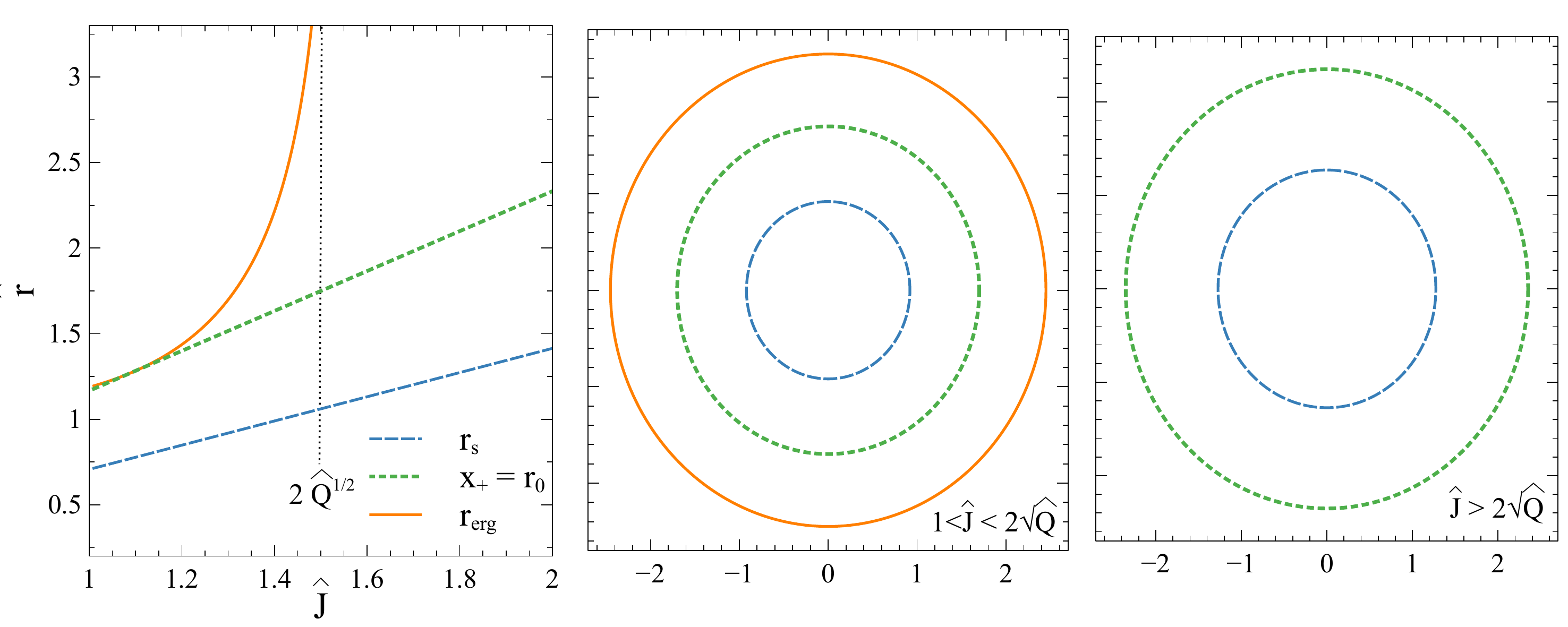}
\caption{\small The WH case: throat and  ergoregion are illustrated. The dashed
line is the curvature singularity, $\hat{r}_s$,  not included in the WH space; the dotted one 
is the throat $r_0=\hat x_+$, and
the solid curve represents the stationary limit, present only if $1<\hat J<2\sqrt{\hat
Q}$. $\hat J>1$ and $\hat Q\leq 1$, to the left $\hat Q=0.6$.}\label{RootsC}
\end{figure}

In summary, for the setting of parameters $\left\{ \hat{J} > 1,\quad \hat{Q} < 1 \right\}$, the metric
(\ref{H_model_sol_BH}) has a structure of $(2+1)$-dimensional stationary cyclic
symmetric AdS wormhole metric with domain $r\in[ r_{0},\infty)$, the wormhole throat
located at $r=r_{0}$, and characterized by the red-shift function $\Phi(r)$, shape function $b(r)$, and
$\omega(r)$ given,  respectively, by (\ref{WH_fun}). Moreover, there exists a
stationary limit if $1< \hat{J} < 2 \sqrt{\hat{Q}}$ i.e. a region where 
$g_{t t}= -N^2(r) - \omega^2(r) =0$. The radios $\hat{r}_0$ and $\hat{r}_{\rm erg}$ are depicted in figure (\ref{RootsC}); the singularity $\hat{r}_s$ does not belong to the WH spacetime.

For self-consistency, the violation of NEC can be checked as well, determining that
the contribution of exotic mater is necessary and is the responsible for keeping open the
WH-throat. Calculating $E_{(\alpha)(\beta)}n^{(\alpha)}n^{(\beta)}$  for the null
vector $ \boldsymbol{n} = (1,1,0)$, it  is obtained that, 

\begin{equation}\label{NECtoL}
8\pi E_{(\alpha)(\beta)}n^{(\alpha)}n^{(\beta)} = 8\pi \left( E_{(0)(0)} +
E_{(1)(1)} \right) = 2\mathcal{H} + 2(2\mathcal{P}\mathcal{H}_{\mathcal{P}} -
\mathcal{H}) = 4\mathcal{P}\mathcal{H}_{\mathcal{P}}, 
\end{equation}
thus, for all $r$ in the WH domain it is fulfilled that,

\begin{equation}
2\pi E_{(\alpha)(\beta)}n^{(\alpha)}n^{(\beta)}  =
\mathcal{P}\mathcal{H}_{\mathcal{P}} = \frac{2\hat J^2 \hat r^2\left(\frac{\hat
J^2}{4\hat{r}^2} +\hat{r}^2-1\right)(1-\hat{Q})}{(\hat{J}^2-2 \hat{r}^2)^3}<0,
\end{equation} 
since, $\hat{Q}<1$, $\hat{N}^2(\hat r)= \left(\frac{\hat{J}^2}{4\hat{r}^2} +
\hat{r}^2-1\right)>0$ and $(\hat{J}^2-2\hat{r}^2)^3<0,$ for all $r$ in the wormhole domain.

\begin{figure}[H]\centering
\includegraphics[scale=0.6]{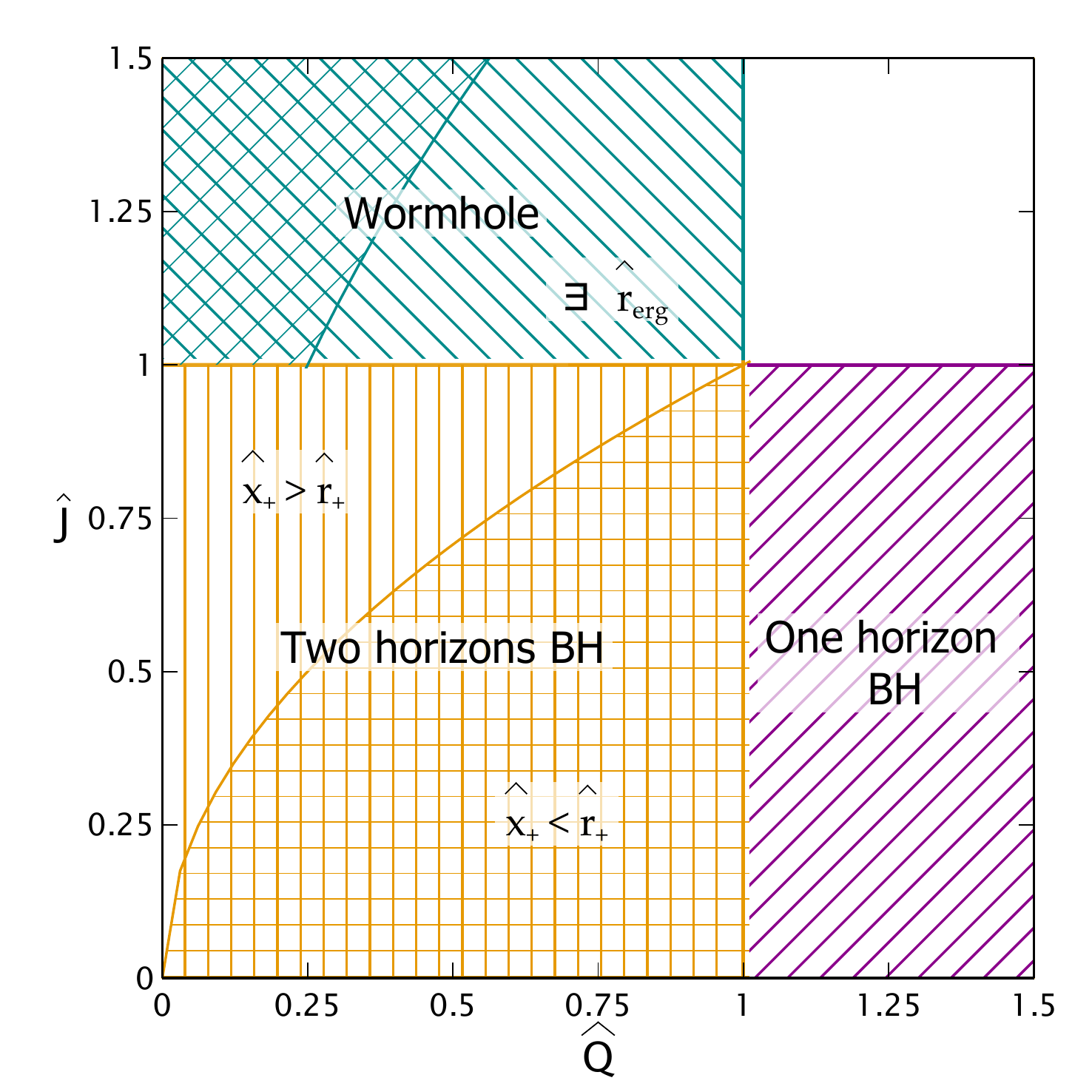}
\caption{\small The parameter space $(\hat{Q}, \hat{J})$ is ullustrated. In the two horizon--BH region the curve  given by  $\hat J=\sqrt{\hat Q}$ signals where the exchange of horizons $\hat x_+$ and $\hat r_+$ occurs. Notice that holding $\hat J\leq 1$ and increasing the electric charge such that $\hat Q>1$, it is possible to go from the two horizon--BH to the one horizon--BH region. In the WH region, observe that if $\hat J< 2\sqrt{\hat Q}$ the WH admits an ergosphere region. }\label{Parameters}
\end{figure}

 In Fig. (\ref{Parameters}) the parameter space $(\hat{Q}, \hat{J})$ is shown. Several remarks arise regarding the ranges of the parameters; for instance, BHs do not admit an arbitrarily large angular momentum, but $\hat{J} \le 1$ or  $J \le l M$, while the electromagnetic parameter $Q$ may be arbitrarily large. The WH does not admit a static limit, since the decreasing  of  $\hat{J}$ turns the WH into a BH. Moreover, it can be seen that starting from the region with two horizons--BH, it is possible to go to the WH region if the angular momentum is increased; or to the region of BH with a single horizon if the electric charge is increased, holding the angular momentum $J$ less than $l M$.

In what follows we back to the notation of parameters without hat.

\section{ The  Magnetic NLED generalization of the Stationary BTZ black hole}

In this section we present the  magnetic NLED generalization of the stationary BTZ
black hole, that is a solution of the field equations in the sub-branch $a\neq0$, $b=0=c$. The field equations are 
explicitly written in the Appendix I. In
this case the only non-null electromagnetic component $P_{(\alpha)(\beta)}$ is given by 

\begin{equation}
P_{(1)(2)} = \frac{a}{N(r)}, 
\end{equation}
that vanishes at spatial infinity. 
The function $\mathcal{H}(\mathcal{P})$, determining the nonlinear electrodynamics source, is given by

\begin{equation}
\mathcal{H}(\mathcal{P}(r)) = - \frac{ a^{s} r^{4} }{2( 4 \Lambda r^{4} + J^{2} )},
\quad \textup{with}\quad s>1 \quad \textup{and}\quad a^{s}\in \mathbb{R},
\end{equation}
while the metric is
\begin{equation}
ds^{2} = - N^{2}(r)dt^{2}  + \frac{(4\Lambda r^{4} + J^{2})^{2} }{ Y^{2}(r) N^{2}(r)} dr^{2} +  
\left[d\phi + \omega(r)  dt\right]^{\!2}, 
\end{equation}

with, 
\begin{eqnarray}\label{2H_model_sol}
N^{2}(r) & = & \left( - M - \Lambda r^{2} + \frac{ J^{2} }{ 4r^{2} }  \right), 
\quad Y^{2}(r) =  \left[J^{4} + 8 J^{2} \Lambda r^{4} + 4(4 \Lambda^{2} - a^{s})r^{8}  
\right], \\
\omega (r) & = & W(r) - \frac{ Y(r)}{2Jr^{2}}, \quad
W(r)  =  \int^{r}  \frac{4[ (4\Lambda^{2}-a^{s})\xi^{4} + J^{2}\Lambda]\xi}{
J\sqrt{  4(4\Lambda^{2}-a^{s})\xi^{8} + 8J^{2}\Lambda \xi^{4}  + J^{4}  } } d\xi.
\end{eqnarray}
The electromagnetic invariants, $\mathcal{P}$ and $F$ are given by (\ref{2invariants}) with $b=0$, 

\begin{eqnarray}
\mathcal{P}(r) & = & \frac{2a^{2}}{ - 4M + \frac{ J^{2} }{ r^{2} } - 4\Lambda r^{2}}, \\
F (r) & = & \mathcal{H}_{\mathcal{P}}^{2}\mathcal{P} =\left(
\frac{\mathcal{H}_{,r}}{\mathcal{P}_{,r}} \right)^{2} \mathcal{P} = \frac{ a^{2s-2}
J^{4} }{2} \frac{  ( - 4\Lambda r^{4} - 4Mr^{2} +  J^{2}  )^{3}  }{ (4\Lambda r^{4}
+ J^{2})^{6}   } r^{6}.
\end{eqnarray}
The invariant $F$  is regular in the whole spacetime except in the region determined
by $r = \left( -\frac{J^{2}}{4\Lambda} \right)^{1/4},$ that corresponds to a
curvature singularity, according to the curvature invariants,

\begin{eqnarray}
R &=& \frac{ \mathcal{B}_{1}(r) }{ ( 4\Lambda r^{4} + J^{2} )^{3} } \label{2R_es}, \\
R_{\alpha\beta}R^{\alpha\beta} &=& \frac{ \mathcal{B}_{2}(r) }{ ( 4\Lambda r^{4} +
J^{2} )^{6} } \label{2Ric_es}, \\
R_{\alpha\beta\mu\nu}R^{\alpha\beta\mu\nu} &=& \frac{ \mathcal{B}_{3}(r) }{
(4\Lambda r^{4} + J^{2})^{6} } \label{2Riem_es},
\end{eqnarray}
where $\mathcal{B}_{n}(r), \quad n= 1,2,3,$ are polynomial functions on $r$, that do not cancel 
the factors $(4 \Lambda r^{4} + J^{2})$ in the denominator, then 
the invariants are finite for all $r$ except at $r = \left( -\frac{J^{2}}{4\Lambda}
\right)^{1/4}.$ Notice that if $\Lambda$ is positive, the curvature invariants do not diverge and the solution is non-singular; this case still may have a BH interpretation with the horizon given by $z_{-}^2= (M- \sqrt{M^2+ \Lambda J^2})/(- 2 \Lambda)$.

The roots of the functions $N^{2}(r)$ and $Y^{2}(r)$ are given, respectively, by

\begin{eqnarray}
&& N^{2}(r) = - M - \Lambda r^{2} + \frac{ J^{2} }{ 4r^{2} }, \quad N^{2}(z_{\pm}) = 0, \quad (z_{\pm})^{2} = \frac{ M \pm \sqrt{ M^{2} + \Lambda J^{2}   } }{ - 2 \Lambda } \label{2BH_rh}
\\
&& Y^{2}(r) = 4(4\Lambda^{2}-a^{s})r^{8} + 8J^{2}\Lambda r^{4}  + J^{4}, \quad Y^{2}(y_{\pm}) = 0,  \quad (y_{\pm})^{2} = \sqrt{ - \frac{ J^{2} }{  2 (2\Lambda \pm \sqrt{a^{s}}) } }.  
\label{2HW_r0}
\end{eqnarray}
It is easy to see that for any $M$, $a$, $s$, $J$, $\Lambda$ $\in \mathbb{R}$ such
that $z_{\pm}$ and $y_{\pm}$ are real numbers, it is fulfilled that,   

\begin{equation}
(z_{-})^{2} \leq r_{s}^{2} \leq (z_{+})^{2} \quad \textup{ and } \quad  (y_{-})^{2}
\leq r_{s}^{2} \leq (y_{+})^{2}, \end{equation}

The quantities $(z_{-})^{2}$ and $(z_{+})^{2}$ are equal only when $\Lambda
= - M^{2}/J^{2}$. In this case we have that $(z_{-})^{2} = (z_{+})^{2} = r_{s}^{2} =
\left( -\frac{J^{2}}{4\Lambda} \right)^{1/2}$. Similarly,  $(y_{-})^{2}$ and
$(y_{+})^{2}$,  are equal to each other only when $a = 0$, implying that $(y_{-})^{2} = (y_{+})^{2} = r_{s}^{2} = \left( -\frac{J^{2}}{4\Lambda} \right)^{1/2}$. 
In what follows we briefly address some particularly interesting cases.

\subsection{BTZ--BH limit}  

\begin{enumerate}

\item {\bf Limit $a\rightarrow0$ with $J\neq0$ }

If $a^{s}=0$ then the electromagnetic field vanishes, $\mathcal{P}=0$. In this case
the function $Y^{2}(r)$ becomes $Y^{2}(r) = (4\Lambda r^{4} + J^{2})^{2}$. Whereas
$W(r)$ becomes $W(r) = \frac{2 \Lambda r^{2} }{ J }$, and then,

\begin{equation}
\omega(r) = W(r) - \frac{ Y(r) }{ 2Jr^{2} } =  \frac{2 \Lambda r^{2} }{ J } - \frac{ 4\Lambda
r^{4} + J^{2} }{ 2Jr^{2} }  = - \frac{ J }{ 2r^{2} }.
\end{equation}
Then the line element (\ref{2H_model_sol}) takes the form,

\begin{equation}
ds^{2} = - \left( - M - \Lambda r^{2} + \frac{ J^{2} }{ 4r^{2} }  \right)dt^{2}  +
\frac{ dr^{2} }{ \left( - M - \Lambda r^{2} + \frac{ J^{2} }{ 4r^{2} }  \right)} +
r^{2}\left(d\phi - \frac{J}{ 2r^{2} } dt\right)^{\!\!\!2},
\end{equation}
that if $M>0$, $\Lambda=-1/l^{2}<0$, corresponds to the stationary BTZ-BH metric.  \\

\item {\bf Limit $J\rightarrow0$ with $a\neq0$ }
In this limit the metric  functions acquire the forms

\begin{equation}
\lim_{J\rightarrow0}\left( W(r) - \frac{ Y(r) }{ 2Jr^{2} } \right) =
\lim_{J\rightarrow0} \frac{1}{J} \left(  \int^{r}  \frac{4[
(4\Lambda^{2}-a^{s})\xi^{4} + J^{2}\Lambda]\xi}{ \sqrt{ 
4(4\Lambda^{2}-a^{s})\xi^{8} + 8J^{2}\Lambda \xi^{4}  + J^{4}  } } d\xi - \frac{
\sqrt{  4(4\Lambda^{2}-a^{s})r^{8} + 8J^{2}\Lambda r^{4}  + J^{4}  }  }{ 2r^{2} }
\right),    
\end{equation}

\begin{equation}
\lim_{J\rightarrow0}\left( W(r) - \frac{ Y(r) }{ 2Jr^{2} } \right) =
\lim_{J\rightarrow0} \frac{1}{J} \left(  \int^{r}  \sqrt{ 4(4\Lambda^{2}-a^{s}) }
\xi  d\xi - \frac{ \sqrt{  4(4\Lambda^{2}-a^{s})  } r^{2}  }{ 2 } \right) = 0.    
\end{equation}
Therefore the line element is

\begin{equation}
ds^{2} = - \left( - M  - \Lambda r^{2}   \right)dt^{2}  + \frac{ 4\Lambda^{2} }{
(4\Lambda^{2}-a^{s})   \left( - M  - \Lambda r^{2}  \right)} dr^{2} 
+  r^{2}d\phi^{2},
\end{equation}
that by renaming $M$ and $\Lambda$ as $\hat{M} = \frac{ (4\Lambda^{2}-a^{s}) M }{ 4\Lambda^{2} }$, $\hat{\Lambda}
= \frac{ (4\Lambda^{2}-a^{s}) \Lambda }{ 4\Lambda^{2} }$ can be written as

\begin{equation}
ds^{2} = - \frac{ 4\Lambda^{2} }{ (4\Lambda^{2}-a^{s})  } \left( - \hat{M}  -
\hat{\Lambda} r^{2}   \right)dt^{2}  + \frac{ dr^{2} }{  \left( - \hat{M}  -
\hat{\Lambda} r^{2}  \right)}  
+  r^{2}d\phi^{2},
\end{equation}
and by rescaling $\frac{ 4\Lambda^{2} }{ (4\Lambda^{2}-a^{s})  } t \rightarrow t$, it becomes

\begin{equation}\label{BJ_0_tilde_generalSol}
ds^{2} = - \left( - \hat{M}  - \hat{\Lambda} r^{2}   \right)dt^{2}  + \frac{ dr^{2}
}{  \left( - \hat{M}  - \hat{\Lambda} r^{2}  \right)}  
+  r^{2}d\phi^{2},
\end{equation}
that is the static BTZ-BH metric.

On the other hand, when $J\rightarrow 0$ the quantities $\mathcal{H}$ and
$\mathcal{P}\mathcal{H}_{\mathcal{P}}$ reduce to 

\begin{equation}
\mathcal{H} =  -\frac{ a^{s} }{8 \Lambda } \quad \textup{ and } \quad
\mathcal{P}\mathcal{H}_{\mathcal{P}} = 0.  
\end{equation}

Therefore the action (\ref{actionL}), with $L =
2\mathcal{P}\mathcal{H}_{\mathcal{P}} - \mathcal{H}$, becomes,

\begin{equation}
S[g_{ab},A_{a}] = \int d^{3}x \frac{\sqrt{-g}}{16\pi} \left( R - 2\Lambda +
\frac{a^{s}}{2\Lambda} \right) = \int d^{3}x  \frac{\sqrt{-g}}{16\pi} \left(R - 
\frac{ 2(4\Lambda^{2} - a^{s})\Lambda }{ 4\Lambda^{2} } \right)    =
\int d^{3}x \frac{\sqrt{-g}}{16\pi}  \left( R - 2\hat{\Lambda}\right). 
\end{equation} 
Confirming that the metric (\ref{BJ_0_tilde_generalSol}) for $\hat{M}>0$, and $\hat{\Lambda} =
-1/\tilde{l}^{2}$,  corresponds to the static BTZ black hole.

\end{enumerate}

Now we just remark some interesting cases of the magnetically charged rotating BTZ-BH generalization without a thorough
analysis. In order that the solution (\ref{2H_model_sol}) be AdS asymptotic, we
should fix $\Lambda=-1/l^{2}$ and $(4\Lambda^{2}-a^{s})>0$. 
This solution has an electromagnetic invariant
$F=\frac{1}{4}f_{\alpha\beta}f^{\alpha\beta}$ well defined in the interval $(r_{s},\infty)$. Regarding the nature of the roots of the metric functions $N(r)$ and $Y(r)$ the interpretation resembles the one of the first solution, analyzed in Section III.

\begin{enumerate}
\item {\bf Case $z_{+}\in \mathbb{R}^{+}$ and $y_{+}\in \mathbb{C}$: BH with  event
horizon at $r_{h} = z_{+}$.} 

If $a^{s}<0$, $M>0$ and  $(M^{2} + \Lambda J^{2})>0$, the metric (\ref{2H_model_sol})
becomes the one of a black hole, with one event horizon at $r_{+} = z_{+} = \sqrt{ { M + \sqrt{ M^{2} + \Lambda J^{2}} }/{ - 2 \Lambda } }$

\item {\bf BH with  event
horizon at $z_{+} = y_{+} \in \mathbb{R}^{+}$, with $a\neq0$, or $\Lambda\neq -
M^{2}/J^{2}$}
In this case
\begin{equation}
\frac{ M + \sqrt{ M^{2} + \Lambda J^{2}   } }{ - 2 \Lambda } = \sqrt{ - \frac{ J^{2}
}{  2 (2\Lambda + \sqrt{a^{s}}) } } \Rightarrow    a^{s} = \left(  -\frac{  2
\Lambda^{2} J^{2}  }{\left( M + \sqrt{ \Lambda^{2} J^{2} + M^{2} } \right)^{2}}  -
2\Lambda \right)^{2}
\end{equation}

\item {\bf Charged rotating wormhole}

If $a^{s}>0$, $4\Lambda^{2}-a^{s}>0$, and $(M^{2} + \Lambda J^{2})<0$, then,
$z_{+}\in \mathbb{C}$ and $y_{+}\in \mathbb{R}$. Then, in this case the solution
(\ref{2H_model_sol}) admits a traversable WH interpretation with wh-throat at
$r=y_{+}.$

\end{enumerate}
\section{Summary}

In this section we present a summary of the obtained results: We presented  two exact solutions of Einstein's  $(2+ 1)$-dimensional gravity coupled to nonlinear electrodynamics (NLED) in Anti-de Sitter spacetime. The solutions are characterized by the mass $M$, angular momentum $J$, cosmological constant or (anti) de Sitter parameter $\Lambda$, and an electromagnetic parameter $Q$. We showed that they represent black holes (BH) or wormholes (WH) depending on the values of the parameters.
The two solutions have as the uncharged  limit the rotating BTZ black hole and the asymptotics at spatial infinity corresponds to the static BTZ-BH,  in contrast to other previously derived charged stationary BTZ black holes within  Maxwell electrodynamics;  therefore the solutions we are presenting are NLED generalizations of the stationary BTZ black hole.  Moreover,  in the limit of vanishing rotation we recover a  NLED static charged BTZ black hole. The derived solutions are singular at certain radius $r_{s} \neq 0$, resembling the ring singularity of the Kerr–Newman spacetime. 

In the first section of the paper we introduced NLED in the hamiltonian formalism in terms of the electromagnetic field $P_{\mu \nu}$ and  determined the classes of electromagnetic fields admitted by the stationary cyclic symmetric (SCS) spacetimes. These classes are:  class 1, with nonvanishing electromagnetic component $P^{t \phi}$  and class 2 with nonvanishing electromagnetic components  $P^{r t }$ and  $P^{r \phi}$.    In  case that the  SCS solution admits a black hole interpretation we showed that the electromagnetic  invariants, $ \mathcal{P}=P_{\mu \nu} P^{\mu \nu}$, in the hamiltonian formalism $H(\mathcal{P})$, and $F=F_{\mu \nu} F^{\mu \nu}$, in the Lagrangian formalism $L(F)$, at least one of them diverges at the BH  horizon; while in the Maxwell case, since $F=P$,  stationary BH solutions with regular electromagnetic   invariants at the horizon are not admissible. Therefore NLED allows a variety of solutions  wider  than  Maxwell electrodynamics.

Then we presented the first solution, a BH that belongs to class 1, with electromagnetic invariant $\mathcal{P}$ being regular, while $F$ diverging at the horizon.  The BH has one event horizon if  $\left\{ J/(lM) \le 1, \quad (M^2- c^{\lambda})/M^2  >1 \right\}$; and  two horizons if   $\left\{ J/(lM) \le 1, \quad (M^2- c^{\lambda})/M^2  \le 1 \right\}$, both with an ergoregion covering the horizon. We briefly addressed the geodesic  motion of test particles, and  solving for the turning points,  we observe regions with negative energy, while the effective potential  barrier can be higher or lower than the BTZ one, but with the same asymptotics.  Furthermore, the solution admits a WH interpretation if the parameters are such that  $\left\{ J/(lM) > 1, \quad (M^2- c^{\lambda})/M^2  < 1 \right\}$ and  it presents an ergoregion if the inequality $ \left\{ 1 <  J/(lM)  < 2 \sqrt{ (M^2- c^{\lambda})  }/M \right\}$  holds, this is illustrated by the parameter space in Fig. (\ref{Parameters}).

Finally we derived a second solution,  characterized by the parameters: mass, $M$, angular momentum, $J$,  electromagnetic charge, $ a^{s}$ and cosmological constant $\Lambda=-1/l^2$. The second solution belongs to subclass 2.a, with electromagnetic invariant $F$ being regular, while $\mathcal{P}$ diverges at the horizon.   We showed that it is a black hole with the appropriate BTZ limits:  in the uncharged case the stationary BTZ--BH is recovered, while if the angular momentum vanishes, the resulting  solution is a charged  static BTZ--BH.  For the second solution we do not perform a thorough analysis as for the first one, but the (polynomial) structure of the metric functions allows us to guess that the cases are very  likely than for the first solution. The difference being in that one solution is electrical while the second one is magnetically charged.
 
Fig. (\ref{Parameters}) suggests that if we consider a two horizon--BH solution with $\hat{J} = 1$ and $\hat{Q} \leq 1$, then by means of a small increase in its angular momentum (for instance by throwing into the BH some particles with the adequate angular momentum) that makes  $\hat{J} > 1$  the BH would be transformed into a WH. Other possibilities of changing the parameters  $\hat{J}$ and $\hat{Q}$ come to mind, that would let us pass from a BH to a WH  or viceversa. It would be interesting to explore such kind of transitions between BH and WH.    

\textbf{Acknowledgments}: N. B. and P. C. acknowledges partial financial support
from CONACYT-Mexico through the project No. 284489. D. M. acknowledges financial
support from CONACYT-Mexico through the PhD scholarship No. 434578. P. C. also thanks
Cinvestav for hospitality.

\section*{Appendix I: Field Equations in the orthonormal basis}

In this appendix we include the explicit form of the NLED field equations. 
For simplicity, one can define an orthonormal frame $\{\theta^{(0)}, \theta^{(1)}, \theta^{(2)}\}$,  
which can be expanded in coordinate frame $\{t, r, \phi\}$ as,

\begin{equation}
\theta^{(0)} = Ndt,  \quad \theta^{(1)} = \frac{dr}{H},  \quad \theta^{(2)} = r(
d\phi + \omega dt),  
\end{equation}
in such a way that in this frame, the metric (\ref{Scyclic}) can be written in
diagonal form, 

\begin{equation}\label{ScyclicMeDiag}
 ds^{2} = g_{(\alpha)(\beta)}\theta^{(\alpha)}\theta^{(\beta)} = -(\theta^{(0)})^{2}
+ (\theta^{(1)})^{2} + (\theta^{(2)})^{2}, 
\end{equation}
where  $g_{(a)(b)} = \eta_{(a)(b)}$  with $\eta_{(a)(b)} = {\rm diag}(-1, 1, 1)$,
and hence in the orthonormal basis some calculations are simplified.  

In order to calculate the non vanishing components of the Einstein tensor in the
orthonormal frame, $G_{(\alpha)(\beta)}$,  we will use the Cartan's structure
equations; $d\theta^{(\alpha)} = - \omega^{(\alpha)}{}_{(\beta)} \wedge 
\theta^{(\beta)}$ and $\frac{1}{2}R^{(\alpha)} {}_{(\beta)(\nu)(\mu)} \theta^{(\nu)}
\wedge\theta^{(\mu)} = d \omega^{(\alpha)} {}_{(\beta)}  +   \omega^{(\alpha)}
{}_{(\nu)} \wedge \omega^{(\nu)}{}_{(\beta)}$, where $\omega^{(\alpha)}$ is the connection 1--form, 
and  the Riemann tensor components relative to the orthonormal frame are given by     

\begin{eqnarray}
&&R_{(0)(1)(0)(1)} = \frac{H}{N}\left[ (HN_{,r})_{\!,r} -\frac{ r^{2}H
}{2N}(\omega_{,r})^{\!2}   \right]- \left(\frac{r\!H\omega_{\!,r}}{2N}  \right)^{2},
\\ 
&&R_{(0)(1)(1)(2)} = \frac{H}{r} \left( \frac{r^{2}H\omega_{,r}}{2N}  
\right)_{\!\!\!,r} + \frac{H^{2}\omega_{,r}}{2N}, \\ 
&&R_{(0)(2)(0)(2)} =  \frac{H^{2}N_{,r}}{rN} +   \left( \frac{r H\omega_{,r}}{2N} 
\right)^{2}, \\
&&R_{(1)(2)(1)(2)} =  - \frac{HH_{,r}}{r} -   \left(\frac{r H\omega_{ ,r}}{2N} 
\right)^{2}.
\end{eqnarray}

The Ricci tensor $\boldsymbol{R} = R_{(\nu)(\mu)} \theta^{(\nu)} \otimes
\theta^{(\mu)}$ and the curvature scalar (or the Ricci scalar) $``R"$ can be defined
by the contractions, so that the Ricci tensor components relative to the orthonormal
frame are calculated as $R_{(\nu)(\mu)} = \eta^{(\alpha)(\beta)}
R_{(\alpha)(\nu)(\beta)(\mu)}$. Whereas the Ricci scalar, $R =
\eta^{(\nu)(\mu)}R_{(\nu)(\mu)}$ is given by,

\begin{equation}
R = 2 \left [-\frac{H}{N}(HN_{,r})_{,r} +  \left( \frac{ rH\omega_{,r} }{2N} 
\right)^{2} - \frac{H^{2} N_{,r}}{rN} - \frac{H H_{,r}}{r}  \right].    
\end{equation}
whereas, the Einstein tensor components in the orthonormal basis are  
defined as $G_{(\nu)(\mu)} = R_{(\nu)(\mu)} - \frac{R}{2}\eta_{(\nu)(\mu)}$, 

\begin{eqnarray}
&&G_{(0)(0)} = -  \left( \frac{rH\omega_{,r}}{2N}  \right)^{2} -  \frac{H
H_{,r}}{r}, \\ 
&&G_{(1)(1)} =  \left (\frac{rH\omega_{,r}}{2N}  \right)^{2} + \frac{H^{2}
N_{,r}}{rN}, \\ 
&&G_{(2)(2)} =  \frac{H}{N}(HN_{,r})_{,r} - \frac{3}{4}
\left(\frac{rH\omega_{,r}}{N}  \right)^{2},\\
&&G_{(0)(2)} =  -\frac{H}{r} \left( \frac{r^{2}H\omega_{,r}}{2N}  \right)_{,r} -
\frac{H^{2}\omega_{,r}}{2N}.
\end{eqnarray}

By  comparison between $\boldsymbol{P}$ written in
the frame 
$\{ dx^{\alpha} \wedge dx^{\beta} \}_{x^{\alpha},x^{\beta} = t,r,\phi}$ and the one
written in  $\{ \theta^{(\alpha)} \wedge   \theta^{(\beta)} \}_{(\alpha),(\beta) =
0,1,2}$,  it  is obtained that,

\begin{equation}
 \boldsymbol{P} = P_{tr} dt \wedge  dr + P_{r\phi} dr \wedge d \phi + P_{t\phi} dt
\wedge  d\phi 
= P_{(0)(1)} \theta^{(0)} \wedge \theta^{(1)} 
 + P_{(0)(2)} \theta^{(0)} \wedge  \theta^{(2)} + P_{(1)(2)} \theta^{(1)} \wedge
\theta^{(2)},   
\end{equation}
from which it can be determined $P_{(a)(b)}$ in terms of $P_{ab}$.
On the other hand, the energy-momentum tensor components in the frame 
$\{ \theta^{(0)},  \theta^{(1)}, \theta^{(2)} \}$ are given by, 

\begin{equation}
4\pi E_{(\alpha)(\beta)} =  \left( 2\mathcal{P}\mathcal{H}_{\mathcal{P}} -
\mathcal{H} \right)\eta_{(\alpha)(\beta)} - \mathcal{H}_{\mathcal{P}}
P_{(\alpha)(\nu)}P_{(\beta)}{}^{(\nu)}.     
\end{equation}
Thus, one finds that the non null energy-momentum tensor components in the
orthonormal frame become,

\begin{eqnarray}
&&4\pi E_{(0)}{}^{(0)} = - \mathcal{H}_{\mathcal{P}}( P_{(0)(1)} P^{(0)(1)}  + 
P_{(0)(2)} P^{(0)(2)}  )  + 2\mathcal{P}\mathcal{H}_{\mathcal{P}} - \mathcal{H},
\quad 4\pi E_{(1)}{}^{(0)} = - \mathcal{H}_{\mathcal{P}} P_{(1)(2)} P^{(0)(2)}, \\ 
&&4\pi E_{(1)}{}^{(1)} = - \mathcal{H}_{\mathcal{P}}( P_{(1)(0)} P^{(1)(0)}  +
P_{(1)(2)} P^{(1)(2)}  ) + 2\mathcal{P}\mathcal{H}_{\mathcal{P}} - \mathcal{H},
\quad 4\pi E_{(2)}{}^{(0)} = - \mathcal{H}_{\mathcal{P}} P_{(2)(1)} P^{(0)(1)}, \\
&& 4\pi E_{(2)}{}^{(2)}= - \mathcal{H}_{\mathcal{P}}( P_{(2)(0)} P^{(2)(0)} +
P_{(2)(1)} P^{(2)(1)}  )  + 2\mathcal{P}\mathcal{H}_{\mathcal{P}} - \mathcal{H},
\quad 4\pi E_{(2)}{}^{(1)} = - \mathcal{H}_{\mathcal{P}} P_{(2)(0)} P^{(1)(0)}. 
\end{eqnarray}

Since $G_{(1)(2)} = 0 = g_{(1)(2)}$, then (via Einstein equations) one arrives at 
$E_{(1)(2)} = 0 \Rightarrow  f_{(2)(0)}f^{(1)(0)} = 0 \Rightarrow bc = 0.$ While,
$G_{(1)(0)} = 0 = g_{(1)(0)}$ together with $bc=0$, implies  $ac=0$.  This
constitutes another way to prove the vanishing conditions $ac=0=bc$, Eq.
(\ref{Theorem1a}).

The field equations of general relativity (with cosmological constant) coupled to
NLED in a stationary and cyclic symmetric (2+1) spacetime, for the branch $c\neq 0$, 
written in the orthonormal frame become,    

\begin{eqnarray}
&& G_{(0)}{}^{(0)} = 8\pi E_{(0)}{}^{(0)} - \Lambda \delta_{(0)}^{(0)}  \quad 
\Rightarrow  \quad \left( \frac{ r H\omega_{,r} }{2N}   \right)^{2} + 
\frac{ (H^{2})_{,r} }{2r} = - 2\mathcal{H}  - \Lambda, \label{field_C_1}\\
&&G_{(1)}{}^{(1)} = 8\pi E_{(1)}{}^{(1)} - \Lambda \delta_{(1)}^{(1)}  \quad
\Rightarrow  \quad  \left( \frac{ r H\omega_{,r} }{2N}   \right)^{2} + 
\frac{ H^{2} N_{,r} }{r N} = 2\left(2\mathcal{P}\mathcal{H}_{\mathcal{P}} -
\mathcal{H}\right) - \Lambda, \label{field_C_2}\\
&&G_{(2)}{}^{(2)} = 8\pi E_{(2)}{}^{(2)} - \Lambda \delta_{(2)}^{(2)}  \quad
\Rightarrow  \quad \frac{ H (H N_{,r} )_{,r} }{N} - \frac{3}{4} \left(  
\frac{ r H\omega_{,r} }{N}   \right)^{2} = -2 \mathcal{H} - \Lambda,
\label{field_C_3}\\
&&G_{(2)}{}^{(0)} = 8\pi E_{(2)}{}^{(0)} - \Lambda \delta_{(2)}^{(0)}  \quad
\Rightarrow  \quad \frac{H}{r} \left(  \frac{ r^{2}H\omega_{,r} }{2N}   \right)_{,r}
+ \frac{ H^{2}\omega_{,r} }{2N} = 0, \label{field_C_4}\\
&&\nabla_{(a)}(P^{(a)(b)}) = 0 = \nabla_{(a)}[ \mathcal{H}_{\mathcal{P}} (
_{\ast}\boldsymbol{P} )^{(a)}  ]  \quad   \Rightarrow   \quad  \mathcal{P} =
-\frac{1}{2} \left( \frac{3c}{ r N \mathcal{H}_{\mathcal{P}} }  \right)^{2}.
\label{field_C_5}
\end{eqnarray}

While in the sub-branch  $\left\{ a \ne 0, \quad b=0=c \right\}$  the field equations, 
in the orthonormal frame, are given by 

\begin{eqnarray}
&& G_{(0)}{}^{(0)} = 8\pi E_{(0)}{}^{(0)} - \Lambda \delta_{(0)}^{(0)}  \quad  \Rightarrow  \quad \left( \frac{ r H\omega_{,r} }{2N}   \right)^{2} + 
\frac{ (H^{2})_{,r} }{2r} =  2\left(2\mathcal{P}\mathcal{H}_{\mathcal{P}} - \mathcal{H}\right) - \Lambda, \label{2field_C_1}\\
&&G_{(1)}{}^{(1)} = 8\pi E_{(1)}{}^{(1)} - \Lambda \delta_{(1)}^{(1)}  \quad \Rightarrow  \quad  \left( \frac{ r H\omega_{,r} }{2N}   \right)^{2} + 
\frac{ H^{2} N_{,r} }{r N} =  - 2\mathcal{H} - \Lambda, \label{2field_C_2}\\
&&G_{(2)}{}^{(2)} = 8\pi E_{(2)}{}^{(2)} - \Lambda \delta_{(2)}^{(2)}  \quad \Rightarrow  \quad \frac{ H (H N_{,r} )_{,r} }{N} - \frac{3}{4} \left(  
\frac{ r H\omega_{,r} }{N}   \right)^{2} = -2 \mathcal{H} - \Lambda, \label{2field_C_3}\\
&&G_{(2)}{}^{(0)} = 8\pi E_{(2)}{}^{(0)} - \Lambda \delta_{(2)}^{(0)}  \quad \Rightarrow  \quad \frac{H}{r} \left(  \frac{ r^{2}H\omega_{,r} }{2N}   \right)_{,r} + \frac{ H^{2}\omega_{,r} }{2N} = 0, \label{field_C_4}\\
&&\nabla_{(a)}(P^{(a)(b)}) = 0 = \nabla_{(a)}[ \mathcal{H}_{\mathcal{P}} ( _{\ast}\boldsymbol{P} )^{(a)}  ]  \quad   \Rightarrow   \quad  \mathcal{P} = \frac{a^{2}}{2N^{2}(r)}. \label{2field_C_5}
\end{eqnarray}

\section*{Appendix II: Connection between the flaring out condition and the
violation of the Null Energy Condition}

Let us consider a null vector in the orthonormal frame, $ \boldsymbol{n} = (1,1,0) =
n^{(\alpha)}e_{(\alpha)}$ with $e_{(\alpha)}$ such that
$\theta^{(\alpha)}e_{(\beta)} = \delta^{(\alpha)}_{(\beta)}$; (we can see that  $
\boldsymbol{n}$ is a null vector since $ ds^{2}( \boldsymbol{n}, \boldsymbol{n}) =
g_{(\alpha)(\beta)}n^{(\alpha)}n^{(\beta)} = 0$), 
then contracting the Einstein tensor with $ \boldsymbol{n}$ we find,

\begin{equation}
G_{(\alpha)(\beta)} n^{(\alpha)} n^{(\beta)} =
\frac{1}{r} \left( 1 - \frac{b(r)}{r}  \right) \Phi_{,r}(r) - \frac{1}{2r^{3}}
\left[  b(r) - r b_{,r}(r)  \right], 
\end{equation}
this equation when evaluated at the throat,  $r=r_{0}$, and using the fact that 
$\Phi(r)$ is continuous and finite everywhere, then $G_{(\alpha)(\beta)}
n^{(\alpha)} n^{(\beta)}|_{r=r_{0}}$ reduces to, 

\begin{eqnarray}
 G_{(\alpha)(\beta)}n^{(\alpha)}n^{(\beta)}  \Big|_{r=r_{0}} = 
-\frac{1}{2r^{3}_{0}} \left[ b( r_{0}) - r_{0}b_{,r}(r_{0})  \right]
=\frac{1}{2r^{2}_{0}} \left[ b_{,r}
(r_{0})- 1  \right]< 0, \quad  {\rm  since } \quad  b_{,r}(r_{0}) < 1.
\end{eqnarray}
Then, by using the Einstein field equations and since $ \boldsymbol{n}$ is a null
vector, one can conclude that
the Null Energy Condition (NEC),  which establishes that
$E_{(\alpha)(\beta)}n^{(\alpha)}n^{(\beta)}\geq0$  for any null vector
$n^{(\alpha)}$, is violated,  since at the wormhole throat $r_{0}$, for $
\boldsymbol{n} = (1,1,0)$, it  is obtained that, 

\begin{equation}
E_{(\alpha)(\beta)}n^{(\alpha)}n^{(\beta)}\!\Big|_{r=r_{0}} < 0.     
\end{equation}
In general relativity, given that  the flaring out condition implies the 
violation of the Null Energy Condition, then the violation of the NEC is
unavoidable for a traversable wormhole.


\end{document}